\documentclass{emulateapj} 

\usepackage{color,epsfig,graphics,graphicx,rotating,comment}
\usepackage[T1]{fontenc}

\definecolor{brown}{rgb}{0.6,0.4,0.2} 
\definecolor{purple}{rgb}{0.5,0,0.5}

\shorttitle{Shocked Molecular Gas in HB\,3} 
\shortauthors{}

\newcommand{\kms}{km\,s$^{-1}$}

\newcommand{\spitzer}{\textit{Spitzer}}

\newcommand{\mic}{$\mu$m} 
\newcommand{\ergsb}{erg\,s$^{-1}$\,cm$^{-2}$\,sr$^{-1}$}

\shorttitle{Shocked H$_2$ and Broad CO in HB\,3}

\begin{document}

\title{Shocked Molecular Hydrogen and Broad CO lines from the Interacting Supernova Remnant HB\,3} 
\author{
J. Rho\altaffilmark{1},
T. H. Jarrett\altaffilmark{2,3}
L. N. Tram\altaffilmark{4,5},
W. Lim\altaffilmark{4},
W. T. Reach\altaffilmark{4},
J. Bieging\altaffilmark{6},
H.-G. Lee\altaffilmark{7},
B.-C. Koo\altaffilmark{8},
B. Whitney\altaffilmark{9}
}
\altaffiltext{1}{SETI Institute, 189 N. Bernardo Ave., Ste. 200, Mountain View, CA 94043; jrho@seti.org}
\altaffiltext{2}{Astronomy Department, University of Capetown, Private Bag X3, Rondebosch 7701, RSA, South Africa }
\altaffiltext{3}{Western Sydney University, Locked Bag 1797, Penrith South DC, NSW 1797, Australia}
\altaffiltext{4}{SOFIA Science Center, USRA,
NASA Ames Research Center, MS211-1, Moffett Field, CA 94043}
\altaffiltext{5}{University of Science and Technology of Hanoi, Vietnam Academy of Science and Technology, 18 Hoang Quoc Viet, Hanoi, Vietnam}
\altaffiltext{6}{Steward Observatory, The University of Arizona, Tucson AZ 85721, USA}
\altaffiltext{7}{Korea Astronomy and Space Science Institute, 776 Daedeok-daero, Yuseong-gu, Daejeon 34055, Republic of Korea}
\altaffiltext{8}{Department of Physics and Astronomy, Seoul National University, Seoul 151-742, Korea}
\altaffiltext{9}{Department of Astronomy, University of Wisconsin, 475 North Charter Street, Madison, WI 53706}
\begin{abstract} 

We present the detections of shocked molecular hydrogen (H$_2$) gas in near- and mid-infrared and broad CO in millimeter from the mixed-morphology supernova remnant (SNR) HB\,3 (G132.7+1.3) using Palomar WIRC, the \spitzer\ GLIMPSE360 and WISE surveys, and HHSMT. Our near-infrared narrow-band filter H$_2$ 2.12 $\mu$m images of HB\,3 show that both \spitzer\ IRAC and WISE 4.6$\mu$m emission originates from shocked H$_2$ gas. The morphology of H$_2$ exhibits thin filamentary structures and a large scale of interaction sites between the HB\,3 and nearby molecular clouds. Half of HB\,3, the southern and eastern shell of the SNR, emits H$_2$ in a shape of a $``$butterfly$"$ or $``$W$"$, indicating the interaction sites between the SNR and dense molecular clouds. Interestingly, the H$_2$ emitting region in the south-east is also co-spatial to the interacting area between HB\,3 and the H~II regions of the W3 complex, where we identified star-forming activity.

We further explore the interaction between HB\,3 and dense molecular clouds with detections of broad CO(3-2) and CO(2-1) molecular lines from the southern and southeastern shell along the H$_2$ emitting region. The widths of the broad lines are 8-20 \kms; the detection of such broad lines is unambiguous, dynamic evidence of the interactions between the SNR and clouds. The CO broad lines are from two branches of the bright, southern H$_2$ shell. We apply the Paris-Durham shock model to the CO line profiles, which infer the shock velocities of 20 - 40 \kms, relatively low densities of 10$^{3-4}$ cm$^{-3}$ and strong ($>$200 $\mu$G) magnetic fields.

%\keywords{infrared:ISM:lines and bands - ISM:individual objects (HB\,3) - supernova remnants}

\end{abstract} 

\section{Introduction}
\label{Sintroduction}

Interstellar shocks are ubiquitous in the interstellar medium (ISM) including supernova blast waves, jets and outflows in protostars, and spiral arm density waves in galaxies. The shocks inject energy and momentum into the surrounding regions, and have a strong impact on the local physical conditions, the gas chemistry, and they regulate dust grains through sputtering and shattering. The interstellar gas dynamics is crucial for understanding the evolution of the ISM. The outflows of protostellar jets are shown to be able to clear material in the core, enough to cause termination of the infall phase, which affects star-formation (SF) efficiency in the cloud \citep{matzner00}. The shocks influence the evolution of molecular clouds by injecting supersonic turbulence, and the resulting star formation efficiency affects the star formation rate in galaxies. Studying Galactic supernova shocks provides a relatively clean example of isolated systems. In contrast, shocks from protostars combine with many other phenomena due to strong UV radiation from the central source and additionally continuous fragmentations of molecular clouds may occur resulting in chains of star forming clusters.

Star-formation activity is frequently found in infrared dark clouds \citep{andre13, konyves10, jackson08}. Turbulence may be a source of initiating star formation through cloud fragmentation \citep{maclow04}. What is the source of turbulence that may induce cloud collapse, such as observed in infrared dark clouds? Whether the observed sub-filaments are a result of the fragmentation process \citep{tafalla15}, or  whether  they first  formed  at the  stagnation  points  of  a turbulent velocity field and have been brought together by gravitational contraction on larger scales \citep{smith16} is unclear. Shock waves are known to inject energy into the ISM. The presence of shocked gas (e.g. SiO) often observed from dense cores may be associated with turbulence.

Supernovae (SNe) play a key role in turbulence enhancement \citep{dib06} and galactic wind acceleration in star-forming galaxies \citep{veilleux05}. Galaxy formation models often include strong stellar feedback that predicts too high star formation rates -- a factor of 100 higher than the values observed \citep{muratov15}. Accounting for proper stellar feedback is essential to model galaxy formation, but the computation itself is challenging. A method of local box simulation with a vertical gravitational potential is used to study how SN feedback in a stratified galactic disc drives turbulence and launches galactic winds \citep{martizzi16}. Although, this method cannot yet produce realistic or physical results, the goal is to better understand disc turbulence and galactic outflow feedback by SNe. This modeling attempt stresses the importance to understanding stellar feedback through SNe. Supernova remnants interacting with molecular clouds provide viable laboratories to study the effects of stellar feedback and the dynamic environment due to the shocks from the powerful explosions.

Signs of SNR interactions with molecular clouds include broad CO lines \citep{vandishoeck93, reach99, rho17}, OH 1720 MHz masers \citep{frail96, yusef-zadeh99, hewitt08}, shocked molecular hydrogen (H$_2$) emission \citep{rho01, neufeld08}, and so-called $``$Mixed-morphology” SNRs with center-filled, thermal X-ray emission through evaporating clouds \citep{rho95, rho98, lazendic06, pannuti14}. The 1720 MHz OH masers were detected from $\sim$20 SNRs \citep{frail96, yusef-zadeh99, hewitt08}, and molecular hydrogen H$_2$ cooling lines, as observed in IC 443 \citep[e.g.][]{burton88, kokusho20}, W44 and W28 \citep{reach05} and other SNRs using \spitzer\ GLIMPSE  data \citep{reach06}. The clearest evidence for interaction between SNRs and molecular clouds is the detection of shock-excited molecular emission.  In the millimeter (mm) and submillimeter (submm), observations provide direct, unambiguous evidence of interaction when a broad ($>$6 \kms) line is detected, attributed to dynamic motion of shocked gas \citep{kilpatrick16}. Molecular broad lines are detected from IC443 \citep{vandishoeck93}, 3C~391 \citep{reach99},  W~44 \citep{wootten77, seta04, reach05, anderl14}, W~28 \citep{arikawa99, reach05}, HB~21 \citep{koo01, shinn10} and W~51C \citep{koo97b}. Broad molecular CO lines are detected from the SNR G357.7+0.3 \citep{rho17} and a few other SNRs \citep{kilpatrick16}.

\begin{figure}[!t]
\includegraphics[scale=0.7,width=9.5truecm]{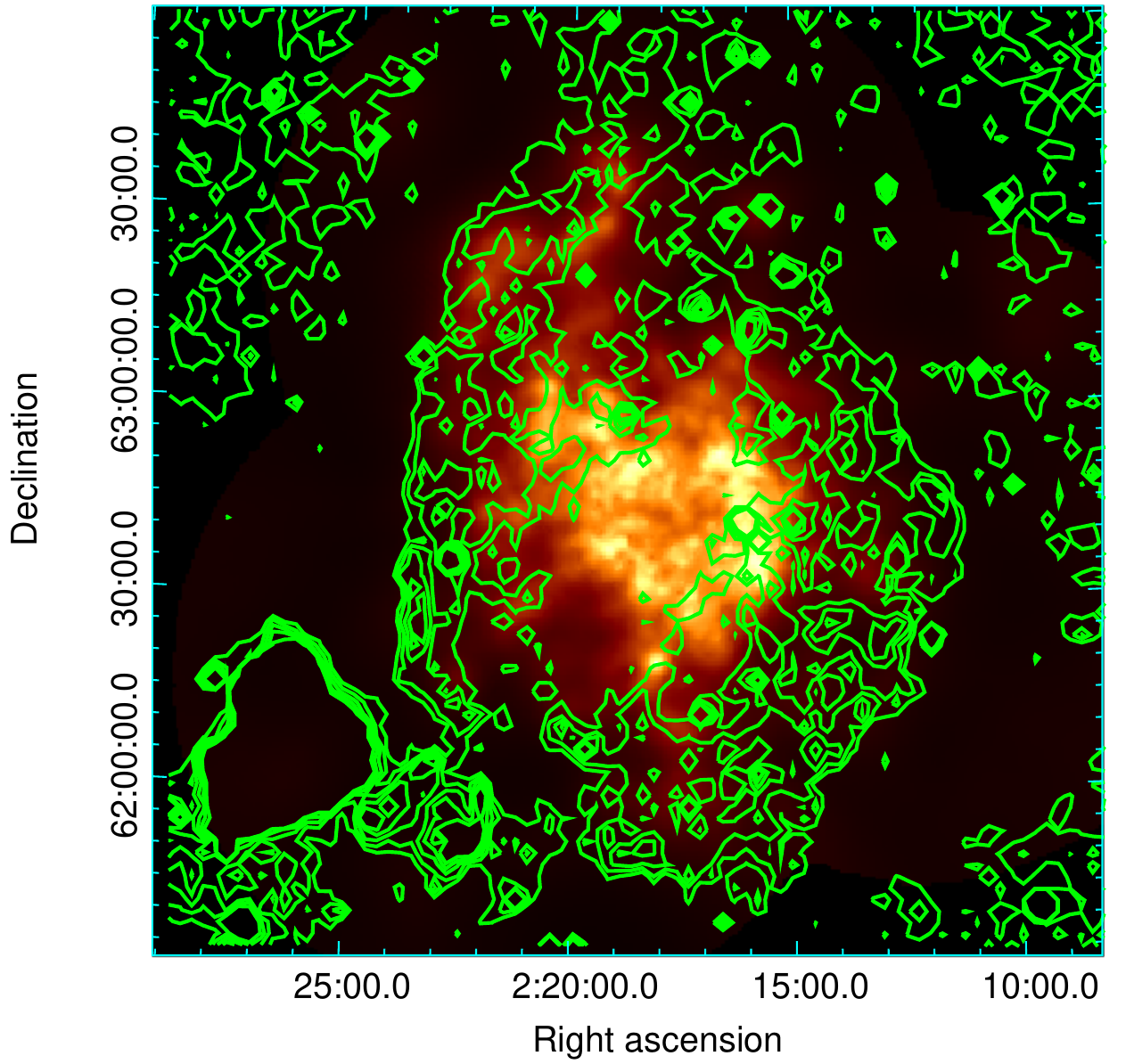}
\caption{ROSAT PSPC X-ray image (0.3 - 2.4 keV) superposed on radio contours of HB\,3 at 92cm (320 MHz) from the WENSS survey {\protect \citep{landecker87}}. All images have J2000 coordinates. The X-ray image has the intensity range of (0.35 - 6.4)$\times$10$^{-3}$ counts s$^{-1}$ arcmin$^{-2}$. The image shows a center-filled X-ray morphology with an inner ring structure. The radio contours are 0.0008, 0.0049, 0.0089, 0.0130, and 0.0170 Jy/beam. are 0.8, 4.9, 8.9, 13, and 17 mJy/beam.}
\label{hb3xrayradio}
\end{figure}

HB\,3 (G132.7+1.3) is a large (80$'$ $\times$95$'$) SNR with a radio spectral index of 0.5 \citep[see Figure \ref{hb3xrayradio};][]{landecker87, leahy85, velusamy74}. A recent estimate of the radio spectrum index is 0.56 \citep{green07}. We assume a distance of 1.95 kpc for HB\,3  \citep{xu06, ruch07}, which is the same as that of the neighboring W3 complex, motivated by GLIMPSE and WISE imaging showing a direct interaction between the SNR HB\,3 and the W3 complex (see Section 4). The optical emission also shows a shell structure, which is roughly circular in outline and brightest at the west and north \citep{vandenbergh73, dodorico77,fesen83, fesen95}.  HEAO-1 and the {\it Einstein} imaging proportional counter (IPC) detected X-ray emission of HB\,3 \citep{leahy85}. The thesis first presented ROSAT PSPC observations by \citet{rho95}, and \cite{rho98} suggested that HB\,3 is a mixed-morphology SNR showing center-filled, thermal X-ray emission within a well-defined shell-like radio morphology as shown in Figure~\ref{hb3xrayradio}. Both of the ROSAT PSPC observations \citep[see][]{rho98, rho95} and ASCA observations \citep{lazendic06, urosevic07} show that the best-fit of X-ray spectra yielded a line-of-sight absorption N$_H =$ (3-9) $\times$$10^{21}$ cm$^{-2}$  and a temperature of $kT =$ 0.15-0.35 keV. The age of HB\,3 estimated from the X-ray temperature is (1.2 - 3.0) $\times$10$^4$ yr (see Section \ref{Sxrays} for details).

%Figure1

%Figure2
\begin{figure*}[!t]
\includegraphics[scale=0.85,width=17.5truecm]{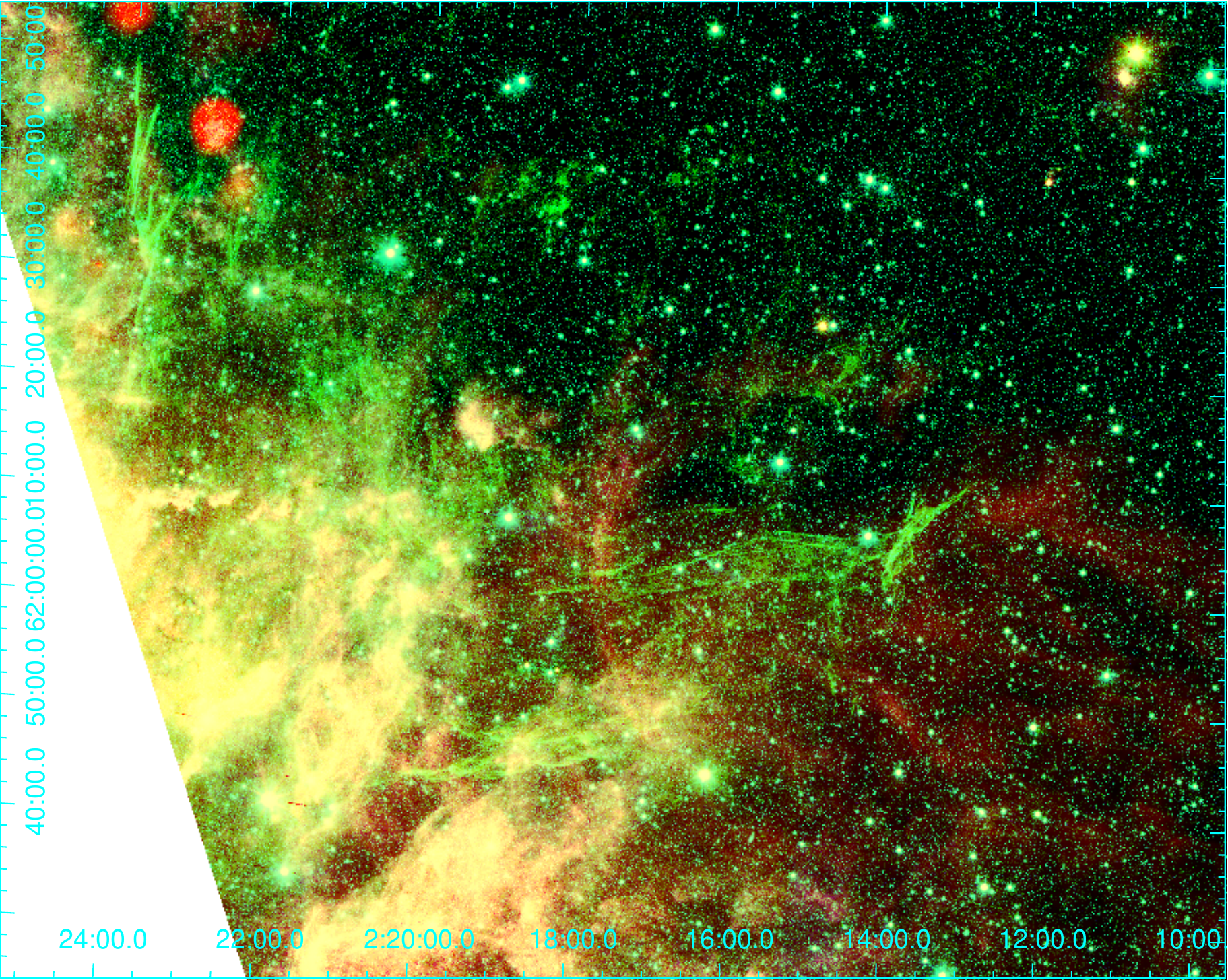}
\caption{Combined \spitzer\ IRAC and WISE mosaic three-color view of HB\,3 as seen in the IRAC 3.6$\mu$m (blue), IRAC 4.5$\mu$m (green) and WISE 12$\mu$m (red). The IRAC 4.5$\mu$m includes shocked molecular hydrogen emission and highlights the regions where the SNR HB\,3 interacts with molecular clouds to the east. Bright H$_2$ filaments appear along the eastern and southern shells, while less filamentary structures are seen along the boundary with the HII region. The eastern and southern filaments along the shell of IRAC-2 show the flux density ranges between 0.3-1.5 MJy sr$^{-1}$ and 0.3-0.95 MJy sr$^{-1}$, respectively. The background has a typical scale of 0.2 MJy sr$^{-1}$.}
\label{hb3spitzeriracall}
\end{figure*}

HB\,3 was suggested to have interactions with nearby clouds as deduced from HI and CO observations \citep{landecker87, routledge91}. Velocity-resolved H~I cloud structures coincide with HB\,3 and an H~I line profile shows a broad line structure between -120 and 20 \kms. However, since the line includes all materials in the line of sight toward HB\,3, the direct interaction is still unclear. There are CO clouds at -42 \kms\ to the south of the SNR \citep{huang86, routledge91} but no broad CO lines were detected. \cite{katagiri16} and \cite{acero16} present bright $^{12}$CO (J=1-0) toward the HB\,3 and the W3 complex, but still without detection of broad CO lines. Moreover, searching for OH masers toward HB\,3 has been confusing due to the bright neighboring HII complex, and \cite{koralesky98} concludes that the masers seen in this direction are not collisionally excited by the shock of HB\,3, and instead, they are most likely associated with the W3 complex. Additionally, \citet{kilpatrick16} made CO observations and suggested that the emission may be associated with the W3 star-forming complex, but not associated with the SNR HB\,3. These prior observational results raises doubt as to whether the SNR HB\,3 is truly interacting with dense molecular clouds. More encouraging, large CO maps (in the millimeter) toward the entire SNR of HB\,3 from the Purple Mountain Observatory show broad CO lines from several positions \citep{zhou16}.

At the highest energy window, $\gamma$-ray emission from HB\,3 was identified using previous missions of COS B \citep{strong77} and EGRET \citep{hartman99}; this was only possible because HB\,3 has a large angular size (and hence fills the beam). {\it Fermi} $\gamma$-ray imaging shows that the $\gamma$-ray emission is spatially correlated with the SNR, originating from the interactions between particles accelerated in the SNR HB\,3 and the decay of neutral pions produced in nucleon-nucleon interactions between accelerated hadrons and interstellar gas \citep{katagiri16,acero16}. {\it Fermi} and HESS observations show $\gamma$-ray emission in many SNRs, but not all \citep{acero16}. It has been suggested that J-type shocks may efficiently accelerate particles to emit $\gamma$-rays \citep{uchiyama10}. The total cosmic-ray spectrum injected into the Galaxy by an individual SNR is estimated by \cite{celli20}.

In this paper,  we present near-IR 2.12$\mu$m narrow-band H$_2$ images along with broad-band \spitzer\ IRAC at 4.5\mic\ (IRAC-2) and WISE 4.6\mic\ ($w2$) images which are interpreted mainly as H$_2$ emission dominated, and detections of broad-line CO emission in the millimeter window. They provide unambiguous evidence of interactions between HB\,3 and molecular clouds, and reveal the large scale structure of interacting sites. Section 3 shows a mosaicked ROSAT X-ray image, and Section 4 presents infrared images of near-IR H$_2$ imaging, together with {\it Spitzer} IRAC and WISE images. Section 5 describes detection of broad-line CO emission using the Heinrich Hertz Submillimeter Telescope (HHSMT or SMT) mm-wave and 12-Meter telescopes, and presents large-scale CO maps covering the southern shell of HB\,3.

\section{Observations}

\begin{table*}
\caption[]{Summary of Millimeter and Infrared Observations }\label{Tobs}
\begin{center}
\begin{tabular}{lllll}
\hline \hline
Date  & Telescope & Lines or filters  \\
1999 Jun 10   & 12-Meter & CO(1-0)  \\
2003 May 17, 18, 19 & HHSMT & CO(3-2), CO(2-1)  \\
2003 June 7, 8  & HHSMT & CO(3-2), CO(2-1) \\
2003 Aug. 9 & Palomar Hale 200" WIRC & H$_2$, continuum \\
2010 June 6, 10, 12, 13  & \spitzer\ IRAC & GLIMPSE360: 3.6 and 4.5$\mu$m \\
2010 Feb. 18 and Aug. 22  & WISE & $w1$ (3.4$\mu$m), $w2$ (4.6$\mu$m), $w3$ (12$\mu$m), $w4$ (22$\mu$m) \\
\hline \hline
\end{tabular}
\end{center}
\renewcommand{\baselinestretch}{0.8}
\end{table*}

\subsection{\spitzer\ GLIMPS360}
HB\,3 is located in the Outer Galaxy at $l$=132.7$^\circ$ and $b$=1.3$^\circ$, one degree away from the Galactic plane. GLIMPSE360 (Whitney et al. 2011; ProgID = 60020) using the \spitzer\ warm mission (at 3.6\mic\ and 4.5\mic) observed the entire Perseus arm ($l$ = 90 -- 190$^{\circ}$) at a distance of 2-4 kpc \citep[see a large-scale gas by][]{alves20}, the Weak Outer Arm (all longitudes) at the distance from the Galactic center, R$_G$ $\sim$ 12 -- 13 kpc, and the Far Outer Galaxy at R$_G$ $>$ 16 kpc. The survey shape followed the mid-plane of the CO/H I warp at R$_G$ = 13 kpc. HB\,3 was observed as part of GLIMPSE360, and the corresponding IRAC 3.6\mic\ (IRAC-1) and 4.5\mic\ (IRAC-2) images are shown in Figure~\ref{hb3spitzeriracall}. 
The IRAC observations used  High-Dynamic-Range (HDR) mode, with short (0.6 sec) and long (12 sec) frames at each position. The array was stepped nearly full-frame in the short direction and nearly 1/3-frame steps in the long direction, yielding a 3-times redundancy and effective exposure time of 32 sec per point.
In this manner we observed every spot in the survey area three times in quick succession. The observation dates of HB\,3 observations are summarized in Table~\ref{Tobs}.

\subsection{WISE observations}

\begin{figure*}[!ht]
\includegraphics[width=18.5truecm]{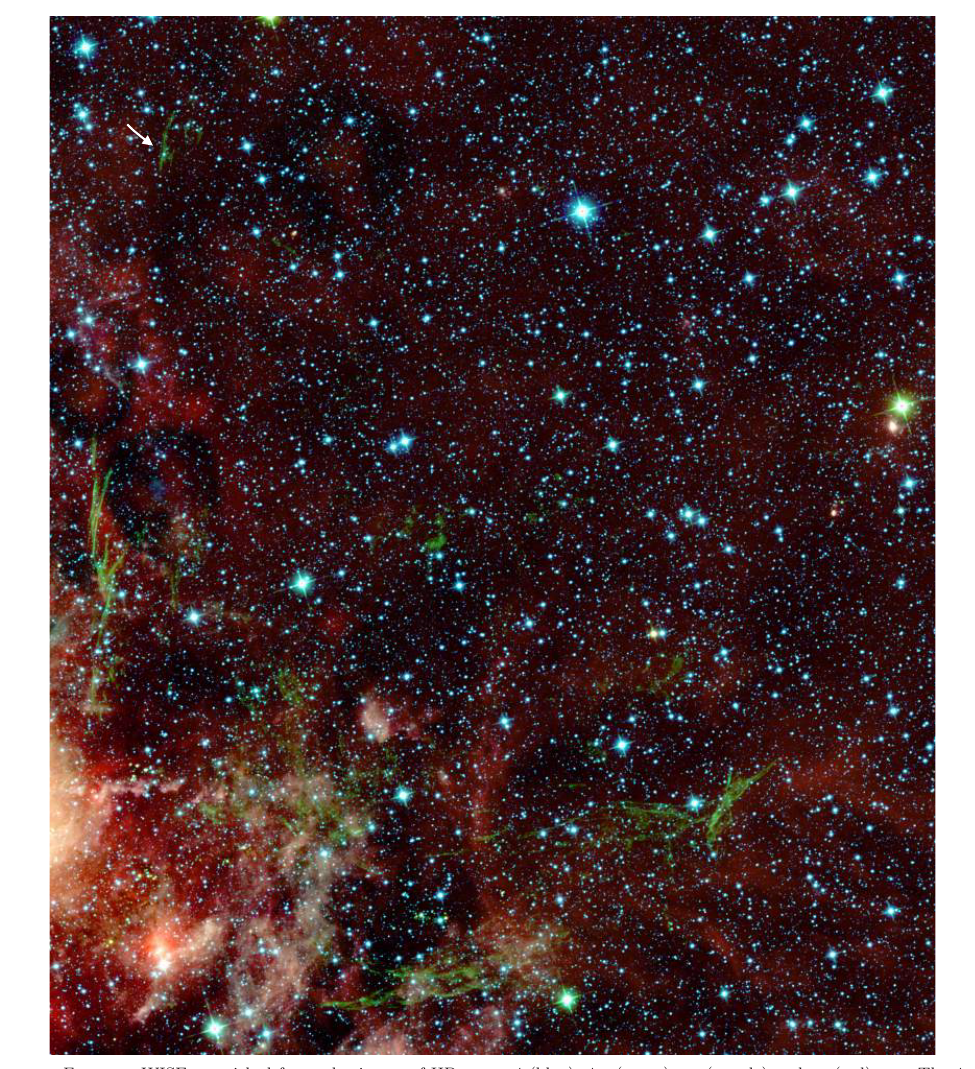}
\caption{WISE four-color mosaic of HB\,3 at 3.4 (in blue), 4.6 (in green), 12 (in orange) and 22 $\mu$m (in red). The 4.6$\mu$m image is dominated by molecular hydrogen and highlights interacting regions between the SNR and molecular clouds. The image is centered on R.A.\ $2^{\rm h} 17^{\rm m} 22.6^{\rm s}$ and Dec.\ $+62^\circ$37$^{\prime}39.9^{\prime \prime}$ (J2000) with a FOV of 103$'$x120$'$. The exact coordinates of the same images are given in Figure \ref{hb3roadmap}.}
\label{hb3wisezoom}
\end{figure*}

\begin{figure}
\includegraphics[width=9.1truecm]{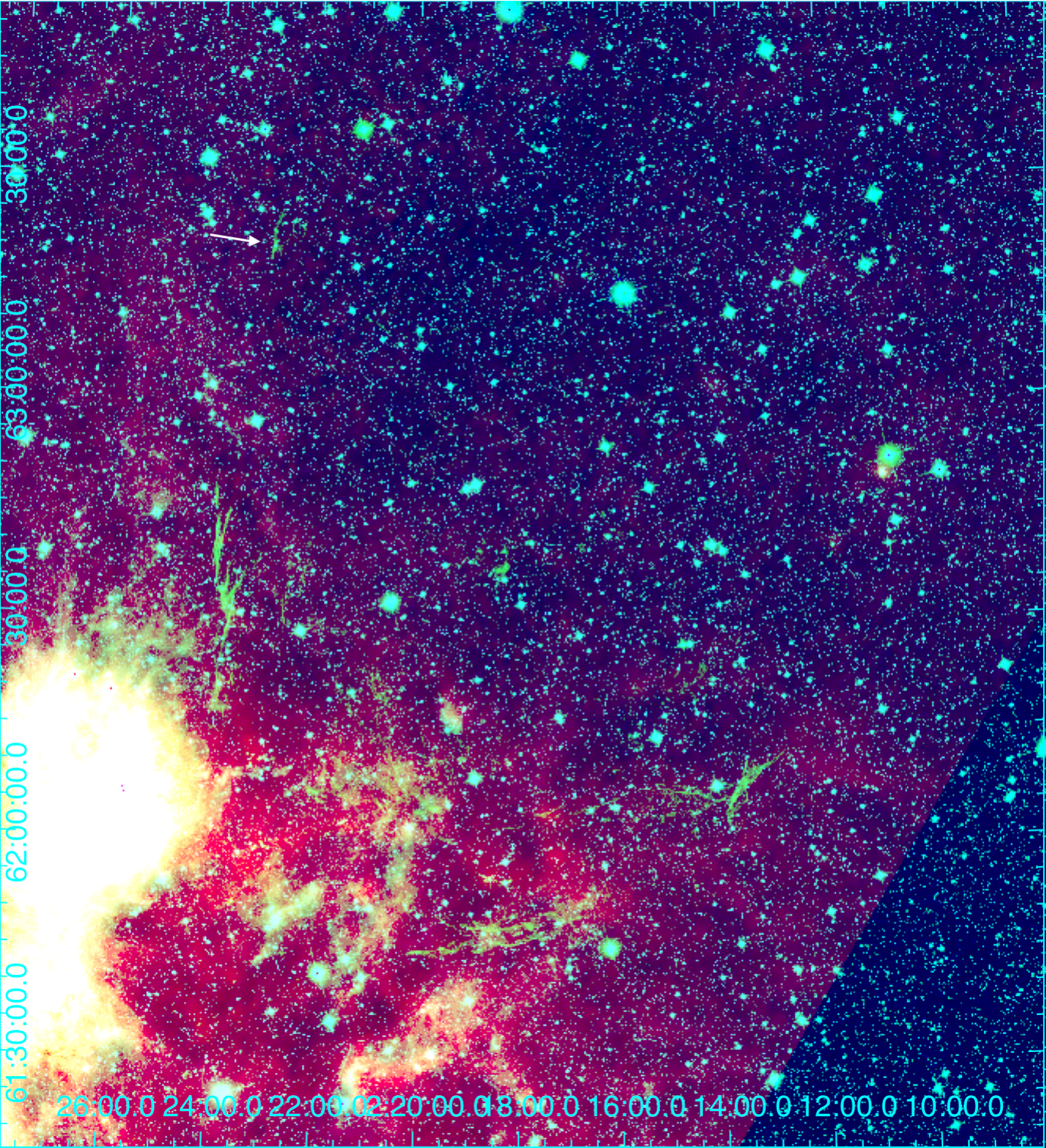}
\caption{WISE-AKARI three-color mosaicked images of HB\,3 at WISE 3.4\mic\ (in blue), WISE 4.6\mic\ (in green), and AKARI 160 $\mu$m (in red). The 4.6\mic\ (in green) filaments are along the 160\mic\ cold clouds from the east to the south of HB\,3, indicating that WISE 4.6\mic\ emission is shocked H$_2$ gas from interacting with cold and dense clouds. The northeastern H$_2$ filament
is marked as an arrow (see Table \ref{Th2pos}.)}
\label{hb3w2a160map}
\end{figure}

Wide-field Infrared Survey Explorer \citep[WISE;][]{wright10} is a whole sky mid-IR survey that uniquely provides data for heavily-obscured SNRs located in the outer Galactic Plane and also beyond Galactic plane. An example of WISE-detected SNR is HB\,3, showing bright filamentary emission in Figures \ref{hb3wisezoom} -- \ref{hb3roadmap}. Unique to the WISE observations and not covered by Spitzer (3.4 and 4.6$\mu$m), it has a broad-band filter centered at 12 $\mu$m ($w3$), and thus encapsulating the ubiquitous 11.3$\mu$m PAH emission arising from the ISM and associated with star formation activity. WISE $w1$ and $w2$ cover 3.4 and 4.6$\mu$m, respectively, sensitive to stellar photospheric emission (continuum), and shock-excited lines, notably in $w2$. The WISE observations of HB\,3  occurred in two passes on 2010, February 18 and August 22.

In order to improve the angular resolution of WISE, we applied high-resolution reconstruction to WISE-detected SNRs \citep{jarrett12, jarrett13}. The nominal spatial resolutions of WISE, $\sim$6$''$ at 3.4, 4.6, and 12\mic\ and 12$''$ at 22$\mu$m \citep{wright10} are two times larger than those of {\it Spitzer} IRAC. The technique takes full advantage of the relatively stable and well-characterized point spread function (PSF) of WISE. For SNRs, this technique brings out the diffuse and filamentary emission distinguishing from point sources. Extragalactic examples are presented in \cite{jarrett13}, demonstrating this technique dramatically improves the extraction of IR structure.

\subsection{Near-IR observations}
\begin{figure*}[!ht]
\includegraphics[scale=0.8,width=18.truecm]{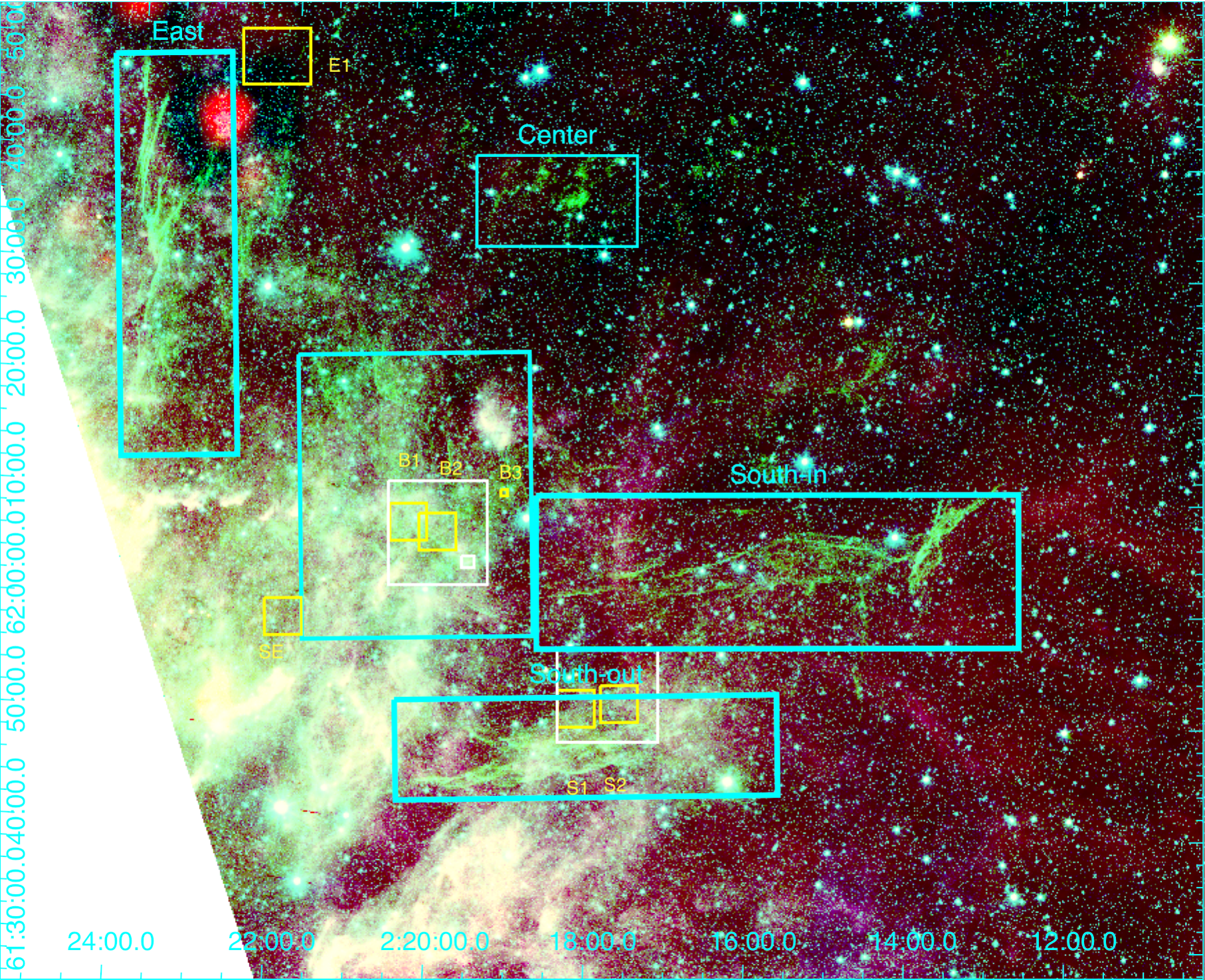}
\caption{Highlighting the most prominent regions of shock-excited molecular hydrogen. Bright shocked H$_2$ emission is de-marked with boxes and labels (in cyan). The white box denotes near-infrared H$_2$ imaging with WIRC (see Fig.~\ref{hb3wirca} for the southern FOV, and see Fig.~\ref{hb3wircb} for the southeastern FOV). The yellow box denotes the FOVs of CO observations (see Fig.~\ref{hb3coh2south} for the southern FOV, and see Fig.~\ref{hb3coh2boundary} for the southeastern FOV).}
\label{hb3roadmap}
\end{figure*}

We observed HB\,3 on 2003, August 8 and 9 with the Wide field InfraRed Camera (WIRC, Wilson et al. 2003) on the Hale 200 inch (5\,m) telescope at Mount Palomar. We took near-infrared WIRC images (with each field of view of $8.7'$) toward the southern shell and southeastern shell, respectively (see Figures \ref{hb3wirca} and \ref{hb3wircb}). The fields of view are marked in Figure \ref{hb3roadmap} (as two white boxes). We observed a set of dithered positions (designed to uniformly cover the SNR), with a corresponding interleaved, dithered observations of an `off' position. A sky reference image was generated from the `off' images, and was subtracted from the `on' images after scaling to match the median sky brightness. This observation and analysis procedure preserved the diffuse emission of the SNR.  In addition to the suite of near-IR filters, we imaged the SNR with a narrow-band filter centered on H$_2$ at 2.12 \mic. The narrow-band filter widths are approximately 0.016 \mic. The exposure time of H$_2$ is a combined 90 sec from three back-to-back images of 30 sec exposure (to reduce the number of saturated stars), with a total on-source integration time of 900 sec for H$_2$. The sensitivity of the narrow-filter H$_2$ images are 1.70$\times$10$^{-6}$ \ergsb\ ($\sim$ 1 Data Number). The detailed data reduction methods are similar to those described in \cite{rho09}; additionally, focal plane spatial distortion was derived from the images and corrected accordingly.

%Figure3

%Figure4

%Figure5

\subsection{Millimeter observations}

We observed the $^{12}$CO J=3-2 (CO\,(3-2), hereafter) line 345.7959 GHz toward a few select positions of the SNR HB\,3 using the HHSMT (or SMT){\footnote[1]{http://www.as.arizona.edu/aro/}} located at Mt. Graham. The positions we observed were selected based on X-ray, radio and previous CO observation to cover the east, south and north parts of HB\,3. The exact positions are summarized in Table \ref{Tpos} and the locations are marked in Figure \ref{hb3roadmap}. The observations took place 2003 May 17-19, and Jun 7-8, before the GLIMPSE and WISE observations were taken in 2010 as listed in Table \ref{Tobs} with the observed positions. We observed $^{12}$CO(2-1) (CO\,(2-1), hereafter) using the acousto-optic spectrometer (AOS) and filter-bank. The AOS spectra  were measured with a 2048-channel, 1 GHz total bandwidth, an effective resolution of 1 MHz. The observations were made with three facility SIS mixer receivers placed at the Nasmyth focus. The beam efficiencies of single-polarization receivers in the frequency bands 210 –- 275 GHz and 320 - 375 GHz are 0.77 for CO(2-1), and 0.48 for CO(3-2), respectively. The telescope beam size is 34$''$ at 217 GHz and 22$''$ at 347 GHz.
A typical RMS noise for CO(3-2) lines we observed is $\sim$0.17 K.

We also observed CO\,(2-1) at 230.53 GHz using the 12-Meter (12\,m) telescope of the National Radio Astronomy Observatory at Kitt Peak on 1999 June 10. The beam size at 230 GHz was 27$''$. The beam efficiency at 230 GHz is 0.52.

\subsection{ X-ray observations of HB\,3}
HB\,3 was observed using the X-ray telescope on ROSAT with the Position Sensitive Proportional Counter (PSPC),  with exposure times 1.3$\times 10^4$ sec. The background corrected counts rate is 2.69$\pm0.05$ cts s$^{-1}$, which is comparable to IPC count rate 2.7 cts s$^{-1}$ (which is not background corrected). X-ray observations of HB\,3 is based on the ROSAT PSPC analysis and initial results reported in the PhD thesis by Rho (1995). The remnant was covered using four pointing observations which pointed at the center, north, south-east and south-west (sequence numbers are 500181n00, 500182n00, 500183n00 and 500184n00, respectively). The images have been mosaicked after exposure and background correction using software by \citep{snowden94}, and  is shown in Figure \ref{hb3xrayradio}.

\section{X-ray Morphology of HB\,3}
\label{Sxrays}

Here we describe ROSAT imaging which provided the detailed X-ray morphology. \citet{lazendic06} and \citet{urosevic07} described ASCA observations which were combined with the ROSAT observations (see below). The mosaicked ROSAT PSPC image is shown in Figure \ref{hb3xrayradio}. The PSPC image shows that the X-ray emission fills the interior, largely circular within the radio shell. The region size emitting its X-ray emission is somewhat smaller than what the radio emission shows; namely, no prominent shell that coincides with the radio shell. One noticeable structure is an X-ray ring inside the radio shell. The radius of the X-ray ring is 16$'$, while the radio shell has a much larger 42$'$ radius. And they are offset,  the center of the X-ray ring is 10$'$ north of the center of the radio shell. A ring feature was resolved within the centrally concentrated emission. Excluding the extension to the north beyond the radio shell that is present in both the radio and X-ray, the remnant is circular with an 80$'$ diameter. The high resolution 408 MHz radio continuum map \citep{landecker87} superposed on the X-ray contours is shown in Figure \ref{hb3xrayradio}. A PSPC hardness map, corresponding to the ratio of a high energy ($\sim$ 0.9-2.2 keV) band map to a low energy band map ($\sim$ 0.5-0.9 keV) shows no spectral variation, indicating there is little variation of column density and temperature \citep{rho98, rho95}. 

The X-ray spectral results yielded a line of sight absorption $N_H$= 4.3($^{+0.25}_{-0.15}$) $\times$10$^{21}$ cm$^{-2}$ and a temperature of $kT$ = 0.33($^{+0.17}_{-0.14}$) keV \citep{rho98}. Assuming HB~3 is in a Sedov phase, the X-ray temperature of 3.8($^{+2.0}_{-1.6}$)$\times$10$^6$ K ($\sim$0.33/0.0862$\times$10$^6$ K) infers a shock velocity ($V_s$) of 522($^{+120}_{-126}$) \kms \citep[$V_s$$\sim$266.6$\times$$\sqrt{T_6}$ \kms\ where $T_6$ is the temperature in the unit of 10$^6$ K, see][]{winkler74, spitzer78}. The angular size of HB\,3 is 65$'$$\times$100$'$, which is equivalent to the radius of 18.5 - 29 pc, which infers an age ($t$) to be 1.8($^{+1.2}_{-0.6}$) $\times$10$^4$ yr ($t$ $\sim$ 41.3$\times$10$^4$ $R_{pc}$/$V_{km/s}$, where $R_{pc}$ is a radius in $pc$ and $V_{km/s}$ is a shock velocity in \kms). The results are consistent with those using the ASCA spectra, which show Si lines at 1.86 keV \citep{lazendic06,urosevic07}, because the X-ray emission of HB~3 is dominated by soft X-ray emission. The mixed-morphology SNRs \citep{rho98} are in a transition phase between Sedov and radiative stages \citep{shelton99} and show recombining plasmas \citep[][references therein]{yamaguchi18}. Thermal conduction and adiabatic expansion play important roles in the center-filled X-ray morphology \citep{gao-yuan19}.
The physical size of HB~3 is 45 $\times$ 53 pc, and the age of HB 3 is (1.2 - 3.0) $\times$10$^4$ yr (see Section \ref{Sxrays}).

%Table2
\begin{table}[!ht]
\caption[]{Observed millimeter HHSMT and near-infrared WIRC positions of HB\,3 }\label{Tpos}
\begin{center}
\begin{tabular}{lll}
\hline \hline
Position &  R.A., Decl. (J2000) & Area (spacing)\\
\hline
Region$^a$ & CO Observations&\\
\hline
S1 (PEAK1)& 02:21:56.10,  +62:04:59.0 & 180$''$$\times$180$''$ (60$''$)\\
S2 (PEAK4)& 02:17:37.75,  +61:51:49.5 & 180$''$$\times$180$''$ (60$''$)\\
B1 (M3, PEAK2)& 02:20:22.56,  +62:07:32.8 & 180$''$$\times$180$''$ (60$''$)\\
B2 (M2, HV)& 02:19:59.95  +62:06:45.7 & 180$''$$\times$180$''$ (60$''$)\\
B3 (M1, PEAK3) &02:19:09.86,  +62:10:25.7 & 30$''$$\times$30$''$ (10$''$)\\
SE (E5) & 02:21:55.41  +61:58:40.9 &420$''$$\times$420$''$ (60$''$)\\
EI  (E1) & 02:22:18.15,  +62:48:40.0 &$360''$ $\times$ $300''$ (60$''$) \\
\hline \hline
WIRC & Near-IR Observations\\
\hline
WIRC1  &2:19:59.636,+62:06:39.07  &9$'$ $\times$ 9$'$ \\
WIRC2  &2:17:46.742,+61:52:34.08 & 9$'$ $\times$ 9$'$ \\
\hline \hline
\end{tabular}
\end{center}
\footnotesize{$^a$The name, $``S"$ indicates the southern shell, and $``B"$ shows $``$Boundary" between the SNR HB\,3 and H~II region of the W3 star-forming complex. $``SE"$ is the SE position of HB\,3, and $``N"$ is the northern position, where a patch of H$_2$ emission is detected. The name in parenthesis is the original name in the HHSMT data}
\renewcommand{\baselinestretch}{0.8}
\end{table}

%Table3
\begin{table*}[!ht]
\caption[]{Bright H$_2$ filaments in HB\,3 }\label{Th2pos}
\begin{center}
\begin{tabular}{lllll}
\hline \hline
Regions &  R.A., Decl. (J2000) & Area ($'$) \\
\hline
Eastern Shell Filament (East) &  2:23:05.03, +62:33:01.56   & 12x31  \\
              & $^a$2:23:57.70, +62:36:09.57   &  --    \\
\hline
Southern Inner Shell (South-In) & 2:13:52.68, +62:04:51.16 & 44x12 \\
                           &$^b$2:14:48.98, +62:06:16.58 & \\
\hline
Southern Outer Shell (South-Out) & 2:18:57.72, +61:47:53.34 & 31x9 \\
                            & $^c$2:18:09.14, +61:45:36.31 & -- \\
\hline
SE Boundary with W3 complex & 2:21:57.49, +62:06:10.04 & 9x8  \\
   (SE-B-H$_2$)     & $^d$2:19:59.60, +62:06:39.10 & --\\
Central H$_2$ position (Center) & 2:18:16.65, +62:36:46.76 & 14x8 \\
$^e$NE shell (NEShell-H$_2$) & 2:22:55:59, +63:22:31.55 & 9x9 \\
\hline
\end{tabular}
\end{center}
{\footnotesize $^{a-d}$ Brightest H$_2$ position for
eastern shell, southern shell 1 and 2, and SE boundary region, respectively (see Figure \ref{hb3roadmap}).}\\
{\footnotesize $^{e}$ A small portion of the north-eastern  shell appears in green (top left) in Figure~\protect\ref{hb3wisezoom}.}
\renewcommand{\baselinestretch}{0.8}
\end{table*}

\section{Infrared Emission}
\subsection{Mid-infrared images using \spitzer\ and WISE}

The \spitzer\ IRAC 4.5$\mu$m image of HB\,3 exhibits thin filamentary structures, particularly bright at the eastern and southern shell structures, as shown in Figure \ref{hb3spitzeriracall}. The filaments coincide with the radio emission of the SNR HB\,3, indicating the filamentary emission belongs to HB\,3 rather than the nearby W3 complex. The combined three-color images of IRAC 4.5$\mu$m with IRAC 3.6$\mu$m and WISE 12$\mu$m images,  Figure \ref{hb3spitzeriracall}, reveals that the filamentary emission (in green) is distinct from the rest of infrared emission, and further confirms that the emission belongs to  HB\,3. The infrared colors from \spitzer\ images (at 3.6, 4.5, 5.8 and 8$\mu$m) were studied by \citet{reach06} and the IRAC 4.5$\mu$m emission (e.g., from SNRs that are interacting with molecular clouds) is shown to be from shocked H$_2$ emission originating from SNRs. We compare the IRAC-2 image with narrow-band H$_2$ images of HB\,3, which strengthens that the emission is from H$_2$ emission (see Section 4.2 for details).

While the filamentary structures are associated with the SNR HB\,3, there is another faint none-filamentary diffuse emission in the area of W\,3 complex near the southeastern part of the SNR HB\,3 (see Figures \ref{hb3spitzeriracall} and \ref{hb3wisezoom}). Two mechanisms have been invoked to account for the H$_2$ line emission: fluorescence induced by ultraviolet pumping \citep{le17, black87}, and radiative cooling following collisional excitation by shock waves \citep{burton89}. The shocked H$_2$ characteristically have broad ($>$ 20 \kms) H$_2$ lines \citep[e.g., see][]{reach19}, while the widths of fluorescent H$_2$ lines are expected to be considerably narrower. The shocked H$_2$ lines have rotational and vibrational excitation temperatures greater than $\sim$2000 K for the H$_2$ lines \citep{burton92}. The faint diffuse emission in the 4.5\mic\ map may relate to the H\,II complex and have a mechanism from the absorption of UV photons or by resonance fluorescence of H$_2$ where the H$_2$ emission is a function of the UV flux \citep{shull82}. The line strengths between 2-1 S(1) and 1-0 S(1) are used to distinguish between the two mechanisms. The ratio is $\sim$ 0.56 for fluorescence of H$_2$ emission and 0.06 - 0.2 for shocked H$_2$ emission \citep{burton89, burton92, le17, black87}. Basically the 1-0 S(1) 2.21\mic\ H$_2$ line is much stronger than the 2-1 S(1) 2.25\mic\ line for shocked gas. The faint diffuse emission may be due to fluorescence induced by ultraviolet pumping from the ionizing star of the H\,II region complex \citep{le17, black87}. 
However, clumpy H$_2$-emitting structures among the diffuse emission are more likely to arise from shock heating since the ambient UV radiation probably does not penetrate the dense clumps.  A classic example of shock heating in complex environments is the
bright H$_2$ emission in the SNR IC\,443 \citep{burton88, rho01, neufeld08}, which show clumpy
and knotty structures and a relatively weak UV radiation field.
Spectroscopic follow-up observations would be required to unambiguously distinguish the two different mechanisms of H$_2$ emission.

Our 2-degree high-resolution reconstructed WISE mosaic covers the entire SNR and nearby W3 complex, Figure \ref{hb3wisezoom}, revealing additional extended H$_2$ emission associated with HB\,3. The $w2$ emission at 4.6$\mu$m exhibits a morphology that is nearly identical to that of \spitzer\ IRAC-2 image, which is to be expected because the bands are very similar, encapsulating the H$_2$ emission line, and the enhanced resolution of WISE is comparable to the beam of IRAC.  The 2-degree mosaic reveals a large scale of interaction site between the SNR HB\,3 and local dense gas from molecular clouds. The WISE and \spitzer\ imaging distinguish between the different interaction sites, notably the boundary with the W3 complex to the east. The H$_2$ emission traces thin filamentary structures, particularly bright emission at the southern and eastern shell. In more detail, this shock-excited emission has a distinctive shape of a $``$butterfly$"$ (or $``$W$"$), which is clearly the interacting sites between the SNR HB\,3 and nearby molecular clouds. Interestingly, it is not just the dense gas of W3 complex, but the interaction zone in the south-east appears to be between HB\,3 and the HII regions of the W3 complex.

The mid-IR imaging highlight the H$_2$ emission at the boundary between the SNR and a nearby HII region of the W3 complex. The north-eastern filament, central, (bright) eastern and southern shell areas are de-marked in Figure \ref{hb3roadmap}, and the positions of bright H$_2$  are summarized in Table \ref{Th2pos}.  About half of the SNR HB\,3 emits H$_2$ emission from east to south, and the emitting area shows a shape of a $``$butterfly$"$ (similar to a rotated letter "W") in projection as shown in Figure \ref{hb3wisezoom}. As noted earlier, this H$_2$ emitting region indicates the interaction between the SNR HB\,3 and nearby molecular clouds. The other half of HB\,3 to the north and west shows no H$_2$ emission, likely because there are no dense gas clouds, or the density of the medium is relatively low overall and ro-vibrational H$_2$ emission is too faint to detect. Figure \ref{hb3w2a160map} shows WISE 4.6\mic\ map superposed on AKARI 160\mic\ cold dust map. The correlation between 4.6\mic\ filaments and cold dust map indicates that WISE 4.6\mic\ emission is shocked H$_2$ gas from interacting with cold and dense clouds. More importantly, the 4.6\mic\ filaments are located along the radio shell of HB\,3 shown in Figure \ref{hb3xrayradio}, indicating the emission is related to the SN shocks.
%Figure6
\begin{figure}
\includegraphics[scale=1,width=7.5truecm]{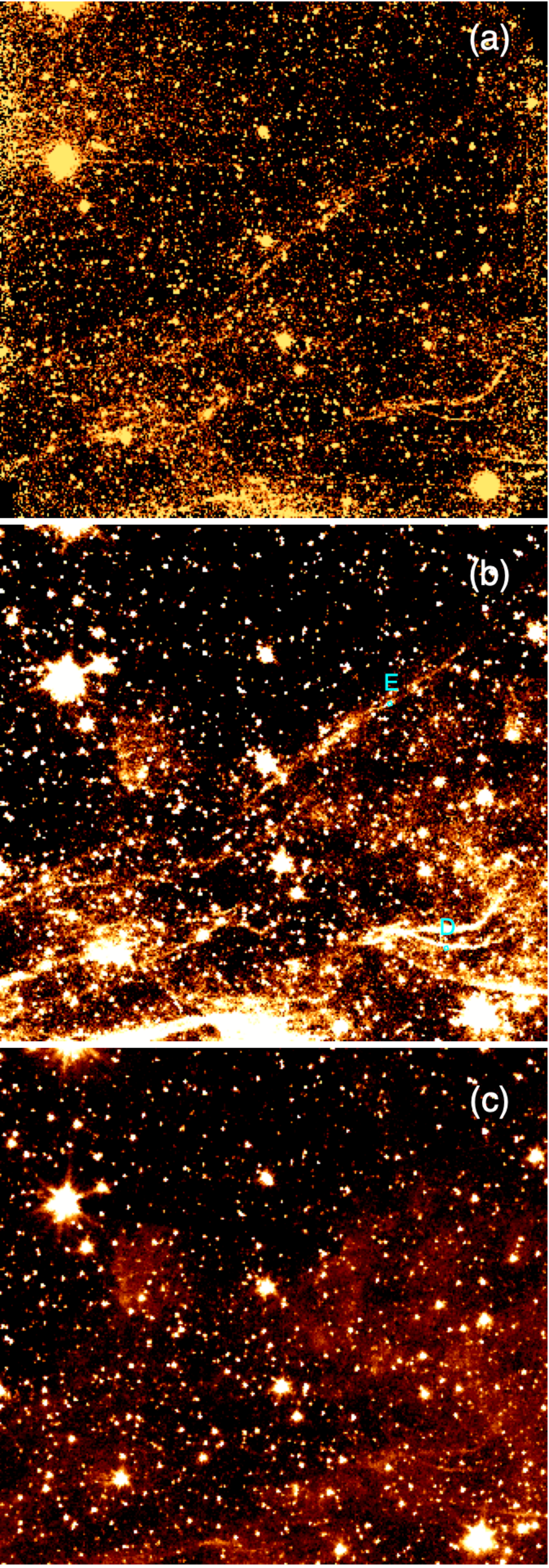}
\caption{Shock-excited molecular hydrogen in the southern shell of the HB\,3 SNR (see Figure \ref{hb3roadmap} and the exact location is marked as a white box). Palomar WIRC narrow-band image (a) and \spitzer\ 4.5$\mu$m (b) and 3.6$\mu$m (c) images centered on R.A.\ $2^{\rm h} 17^{\rm m} 46.7^{\rm s}$ and Dec.\ $+61^\circ$52$^{\prime}34.1^{\prime \prime}$ (J2000) with a FOV of 8.8$'$x8.4$'$. The diffuse emission in the WIRC H$_2$ 2.12$\mu$m image ranges 0.179 - 1.3 MJy sr$^{-1}$ ($\sim$1.91-13.9 $\times$10$^{-6}$ \ergsb). The emission in the \spitzer\ 4.5 and 3.6$\mu$m images ranges 0.3-1.4 (typical background is 0.2) MJy sr$^{-1}$ and 0.44-0.83 (0.35) MJy sr$^{-1}$,  respectively. The filament positions of D and E (see Table \ref{Th2spitzerflux}) are marked on the image (b).}
\label{hb3wirca}
\end{figure}

We provide a finding chart, Figure \ref{hb3roadmap}, of the locations of the IR emission. The eastern filaments, Figures \ref{hb3spitzeriracall} and \ref{hb3wisezoom}, have very sharp edges which appears within the box marked as $``$East" in Figure \ref{hb3roadmap}. Along the Southern shell we see two branches of filaments: one is the outer southern shell and the other is the inner southern shell, marked as $``$South-out" and $``$South-in", respectively. The locations of the IR emission associated with the SNR and the FOVs of the H$_2$ and millimeter CO observations will be discussed in Sections 4.2 and 5, respectively. We provide a finding chart, Figure \ref{hb3roadmap}, of the locations of the IR emission, and the FOVs of the H$_2$ (in Section 4.2) and millimeter CO observations (see Section 5).

The range in morphology is likely due to geometric viewing effects. Similar to the Cygnus Loop \citep[G74.0-8.5][]{hester94}, the mid-IR mosaic of HB\,3 shows a network of filaments. For example, the brightest H$_2$ emission of HB\,3 exhibits very smooth filamentary emission, as if it is from a gently rippled sheet as demonstrated by \citet{hester94} in the Cygnus Loop. The bright emission of the eastern filaments are likely viewed edge-on (in tangential direction relative to the shock). The smoothness of the emission indicates that the pre-shock medium is homogeneous. This morphology of H$_2$ emission in HB\,3 is in stark contrast to the knotty H$_2$ structures seen in IC\,443. From early work in 1980's and 90's, IC\,443 became the prototype mixed-morphology SNR, revealed with near-infrared narrow-filter H$_2$ imaging \citep{burton88, richter95} and 2MASS broad-band imaging \citep{rho03}. Another prominent Galactic SNR is W44, which also shows mostly knotty structures of H$_2$ emission, although faint smooth filamentary structures are also seen \citep{reach05, reach06}. The H$_2$ knotty emission is much brighter than that of whispy filamentary structures for many cases in IC\,443 and W44.  On the other hand, the southern H$_2$ emission in HB\,3 is remarkably similar to the southern part of H$_2$ emission in W44. We note that HB\,3 is special because the remnant shows not only interactions with molecular clouds but also interactions with the W3 HII region complex. A few SNRs have nearby HII regions, for example W44 and W28 \citep{reach05, rho94}, but the HII regions are in projection in the line of sight, and there has not been any cases in which direct interaction with HII regions has been conclusively demonstrated. HB\,3 is therefore a very interesting mix-morphology SNR case.

The mid-IR \spitzer\ IRAC-2 and WISE-$w2$ images reveal that the SNR HB\,3 is interacting with the W3 HII complex. The shocked H$_2$ emission appears at the boundary between them and has a filamentary morphology with clumpy and compressed structures at some locations, indicating shocks. The nature of the shock is still to be determined. They may be from C- or J-shocks, with compression tempered by the underlying magnetic field. Follow-up infrared spectroscopy to detect rotational and vibrational H$_2$ lines will help to distinguish the nature of the shock as it was done for other SNRs which are interacting with clouds \citep{hewitt08} and G357.7+0.3 \citep{rho17}.

In summary, the molecular H$_2$ imaging impressively reveals the boundaries between the SNR HB\,3 and the HII region of the W3 complex, as well as the shocked SN shell itself.

%Figure7 
\begin{figure*}
\includegraphics[scale=1,width=16truecm]{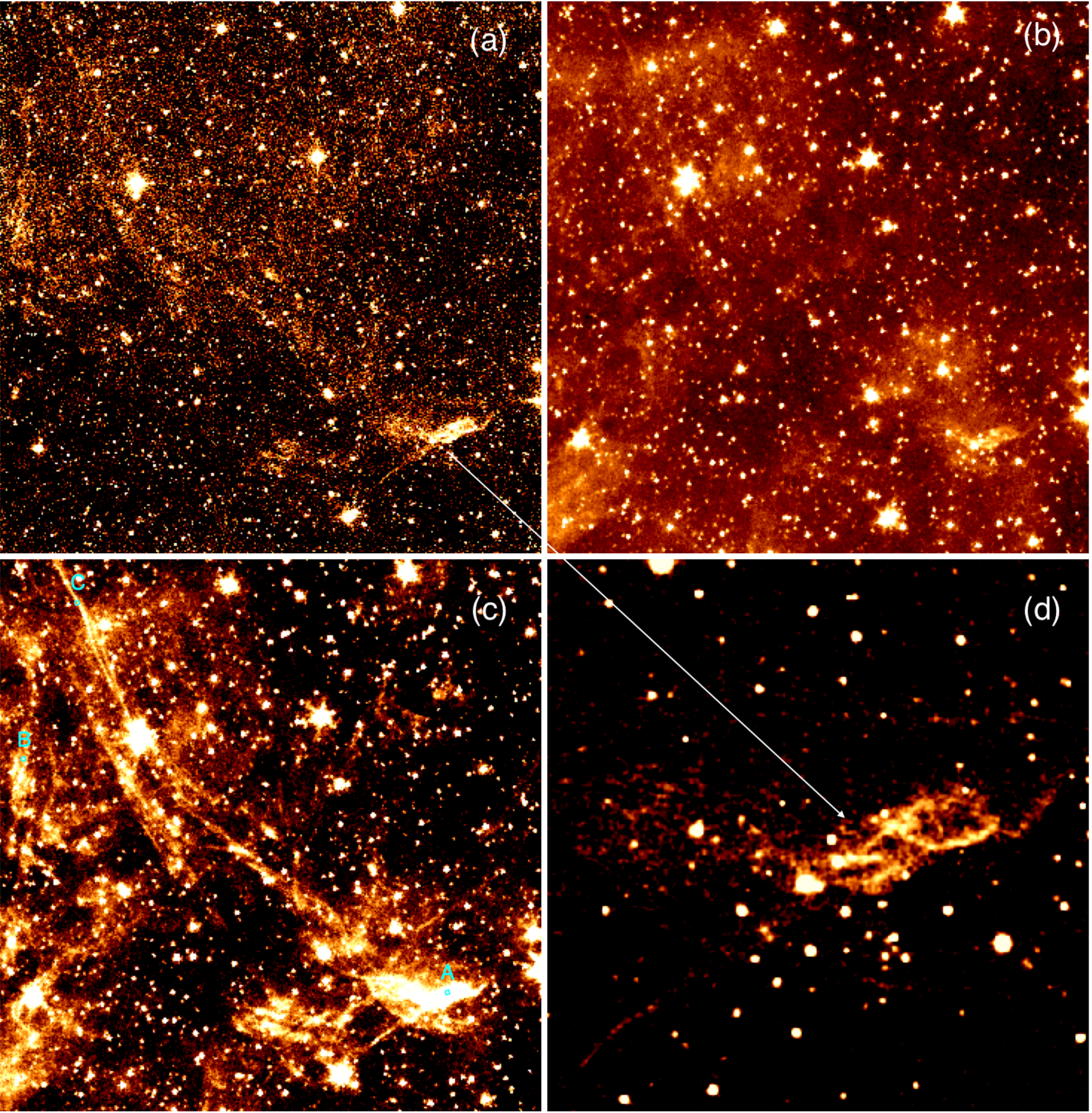}
\caption{Shock-excited molecular hydrogen in the south-east region of the HB\,3 SNR (see Figure \ref{hb3roadmap} and the exact location is marked as a white box). The diffuse emission in the WIRC H$_2$ 2.12$\mu$m image $(a)$ ranges 0.179 - 2.4 MJy sr$^{-1}$ ($\sim$1.91-26.7 $\times$10$^{-6}$ \ergsb). The emission in \spitzer\ 3.6 and 4.5$\mu$m imaging, upper right and lower left $(b, c)$ respectively, ranges 0.41-0.91 (typical background is 0.35) MJy sr$^{-1}$ and  0.25-1.28 (0.2) MJy sr$^{-1}$, respectively. The images {\it (a, b, and c)} are centered on R.A.\ $2^{\rm h} 19^{\rm m} 59.6^{\rm s}$ and Dec.\ $+62^\circ$06$^{\prime}39.1^{\prime \prime}$ (J2000) with a FOV of 8.8$'$x9.2$'$. The narrow-filter WIRC image $(d)$ zoomed on the brightest emission shows complex shock structures, and is centered on R.A.\ $2^{\rm h} 19^{\rm m} 59.6^{\rm s}$ and Dec.\ $+62^\circ$06$^{\prime}39.1^{\prime \prime}$ (J2000) with a FOV of 78$"$x73$"$. The filament positions of A, B, and C on the image (c) (see Table \ref{Th2spitzerflux}) are marked.}
\label{hb3wircb}
\end{figure*}

\subsection{Near-Infrared Palomar WIRC imaging}

We show that the morphology of the \spitzer\ IRAC-2 and WISE $w2$ emission is identical to that in narrow-band H$_2$ imaging as shown in Figures~\ref{hb3wirca} and \ref{hb3wircb}. The WIRC narrow-band H$_2$ 2.12$\mu$m images cover parts of the south and southeast regions. The exact regions observed with WIRC are marked in Figure \ref{hb3roadmap}. The \spitzer\ IRAC images have a 2.5$''$ spatial resolution, while the Palomar WIRC images have a seeing limited spatial resolution with a point-spread-function of 0.6$''$. Figures \ref{hb3wirca} and \ref{hb3wircb} show that the WIRC narrow-filter H$_2$ images are almost identical to those of 4.5$\mu$m \spitzer\ images, revealing very thin filamentary structures, to the resolution limit of our Palomar WIRC imaging (1$''$ = 0.0095 pc in HB\,3). This demonstrates that 4.5$\mu$m \spitzer\ imaging traces H$_2$ emission, as, for example, it has been demonstrated in the case of W44 \citep{reach06}. Figure \ref{hb3wircb} shows filaments from the south-east of HB\,3 and at the bottom right corner of the image shows H$_2$ emission with clumpy or knotty structures. The knotty H$_2$ structures are shown in the zoomed panel in Figure \ref{hb3wircb}d. This region is where the SNR interacts with the W3 complex, but H$_2$ emission still shows nice shell structures. The spatially high resolution H$_2$ imaging with WIRC further confirms that the emission of \spitzer\ IRAC-2 and WISE-$w2$ originates from shocked H$_2$ emission from interaction of the SNR HB\,3 with dense molecular gas.

%Figure8 
\begin{figure}
\includegraphics[scale=0.7,width=8.4truecm,angle=0]{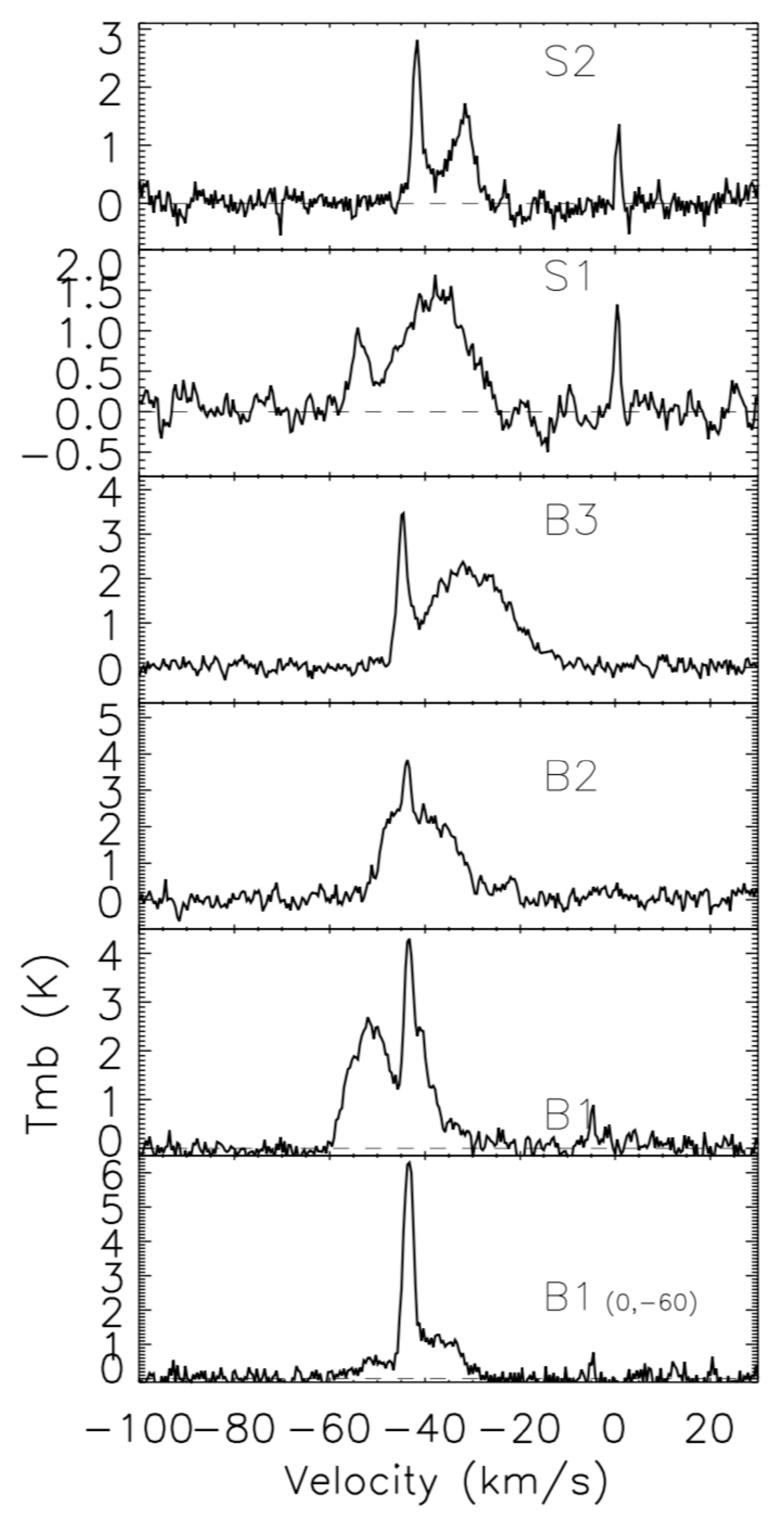}
\caption{HHSMT CO(3-2) spectra toward a few positions of HB\,3. The unshocked, cold clouds are at V$_{LSR}$ $\sim$ -43 \kms (where $LSR$ is the Local Standard of Rest). The order of the spectrum is from west to east. $``$S" indicates the southern shell, and $``$B" indicates the boundary positions between the SNR HB\,3 and the H~II region W3 complex. The CO GRID spectra and H$_2$ images of some of these positions are shown in Figures \ref{hb3coh2south}  and \ref{hb3coh2boundary}, and are listed in Table \ref{Tpos}. The bottom panel of $``$B1(0,-60)" indicates the offset in arcsec from B1 position. The Gaussian fit results are summarized in Table \ref{TCOlines}. Each spectrum shows a CO broad line with a range line widths of 7-21 \kms, and the preshock gas widths at 41 - 44 \kms. The central velocity of shocked gas is at -32, -43, and -47 \kms,  S2, B2, and B1, respectively.}
\label{hb3allcospec}
\end{figure}

%Table4
\begin{table*}
\caption[]{Flux comparison between WIRC 2.1$\mu$m and \spitzer\ 4.5 and 3.6$\mu$m in HB\,3}
\begin{center}
\begin{tabular}{l|l|l|l|l|l}
\hline \hline
Position & WIRC H$_2$ & \spitzer\  &\spitzer\ \\
R.A., Decl. (J2000)                     & 2.1$\mu$m  & 4.5$\mu$m  & 3.6$\mu$m & IRAC-2/H$_2$ & IRAC-2/IRAC-1\\
                 & MJy/sr & MJy/sr &  MJy/sr\\
                 & (\ergsb)  &  & \\
\hline
A; 02:19:35.00, +62:04:08.86 & 1.71$\pm$0.04 (1.83$\pm$0.04 E-5) & 1.72$\pm$0.44 & 0.94$\pm$0.44 & 1.01$\pm$0.25 & 1.83$\pm$0.78 \\
B; 02:20:33.09, +62:07:30.98 & 0.66$\pm$0.03 (7.1$\pm$0.24 E-6) & 0.58$\pm$0.25 & 0.44$\pm$0.22 & 0.88$\pm$0.38  & 1.32$\pm$0.87\\
C; 02:20:26.72, +62:09:57.92 & 0.65$\pm$0.03 (7.00$\pm$0.24 E-6) & 0.56$\pm$0.25 & 0.44$\pm$0.22 & 0.86$\pm$0.39 & 1.27$\pm$0.85\\
D; 02:17:20.56, +61:49:42.75  & 0.40$\pm$0.02 (4.30$\pm$0.19 E-6) &0.54$\pm$0.23 & 0.45$\pm$0.21 & 1.35$\pm$0.58 & 1.20$\pm$0.75\\
E; 02:18:09.95, +61:49:26.92  & 0.40$\pm$0.02 (3.28$\pm$0.17 E-6) & 0.37$\pm$0.19  & 0.38$\pm$0.19  & 1.19$\pm$0.62 & 0.97$\pm$0.69\\
\hline
\hline
\end{tabular}
\end{center}
\renewcommand{\baselinestretch}{0.8}
\label{Th2spitzerflux}
\end{table*}

%Figure9 
\begin{figure*}
\includegraphics[scale=0.7,width=8.8truecm,angle=0]{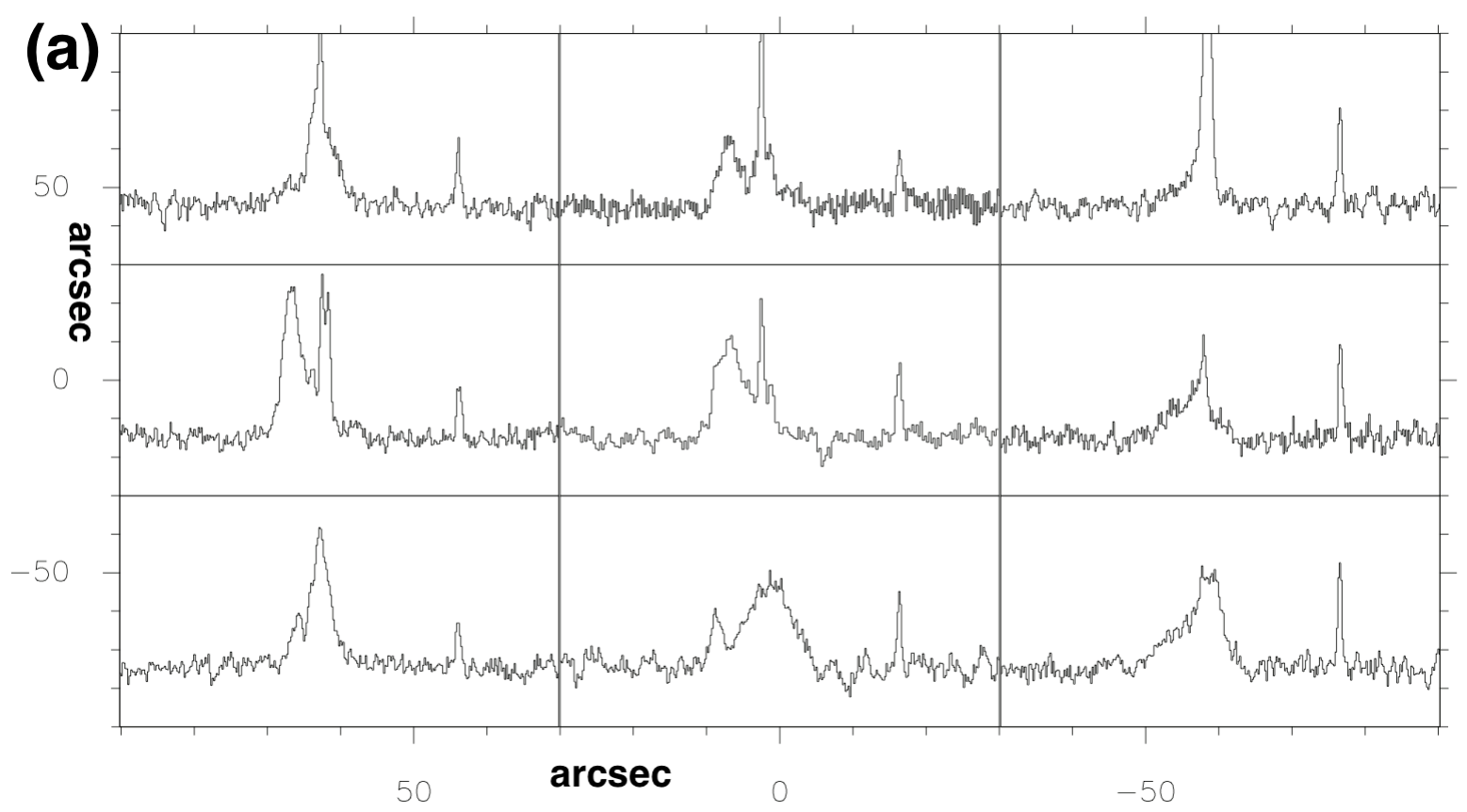}
\includegraphics[scale=0.8,width=8.8truecm,angle=0]{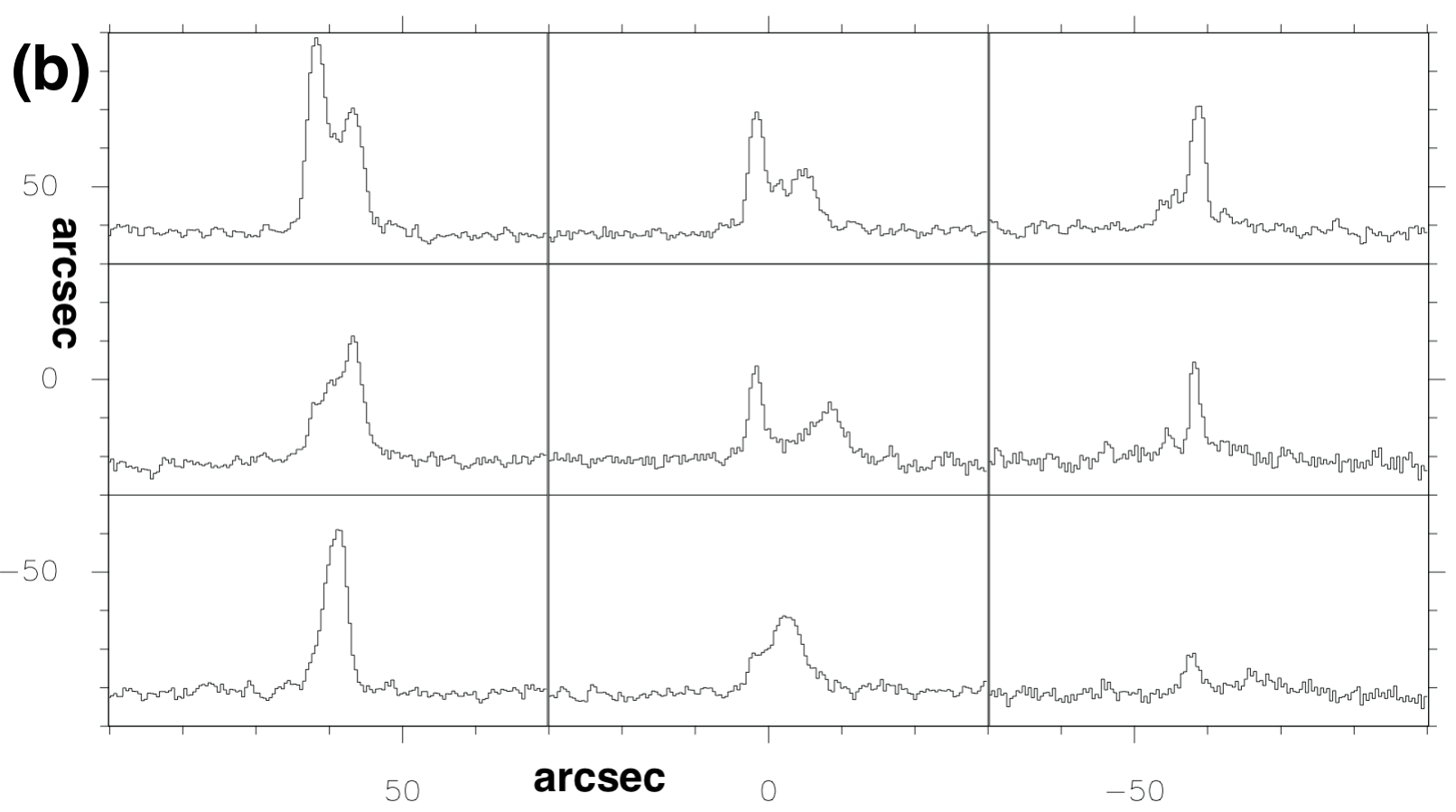}
\begin{center}\includegraphics[scale=1,width=10truecm,angle=0]{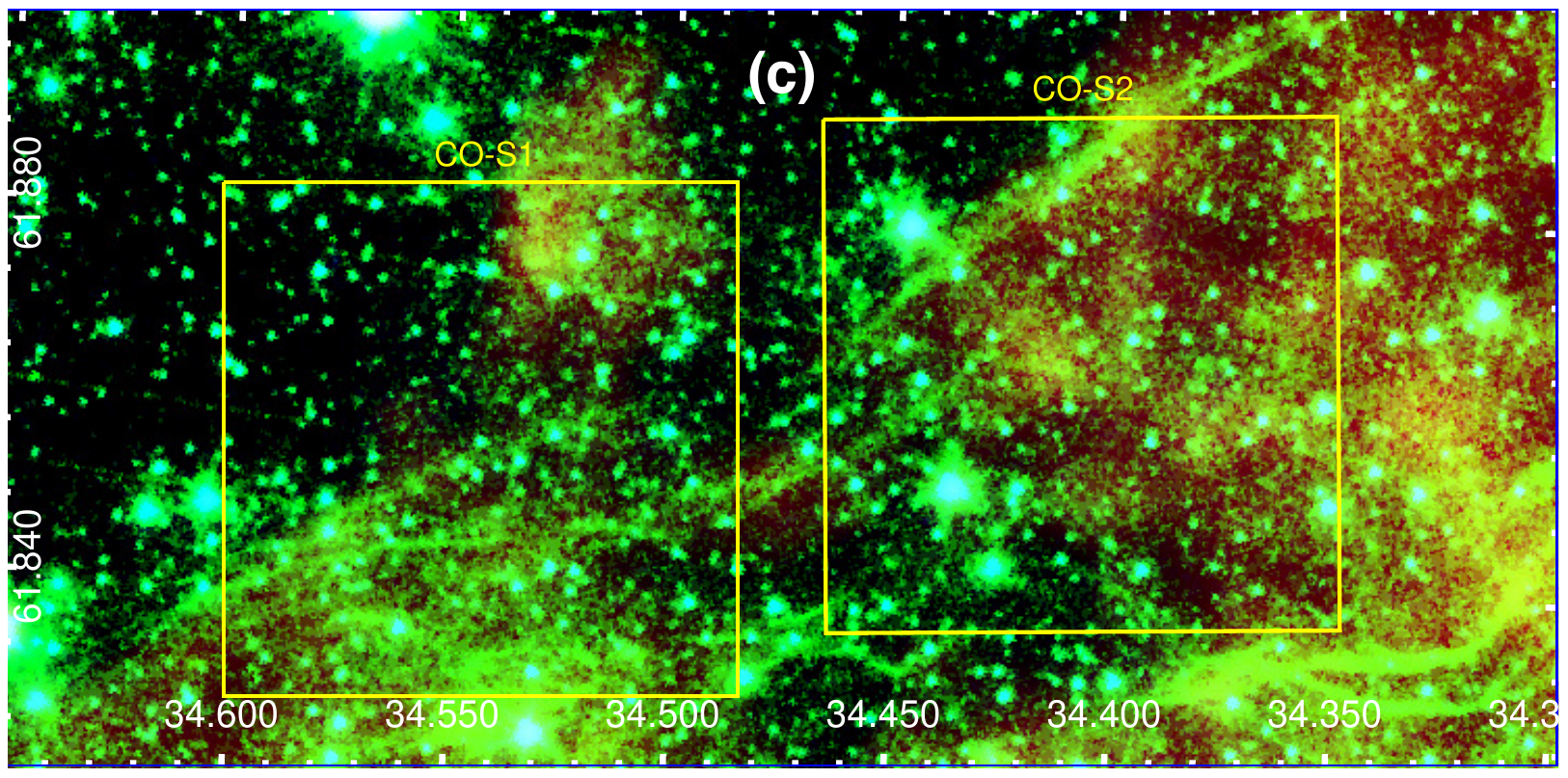}
\end{center}
\caption{Spatial distribution of molecular CO toward the southern shell (the exact locations are marked as yellow boxes on a large map of HB\,3 in Figure \ref{hb3roadmap} in sexagesimal coordinate.) GRID spectra (arcsec in x- and y-axis) of CO(3-2) are toward S1 $(a)$ and S2 $(b)$ regions. The covered FOVs of the GRID spectra are marked on the \spitzer\ 4.5$\mu$m (in green) and WISE (in red) images $(c)$. For individual CO spectrum in CO-S1 region $(a)$, the x-axis ranges in velocity from -100 to +30 \kms\ and the y-axis from -1 K to 3 K. For individual spectrum in CO-S2 region $(b)$, x-axis is a velocity from -70 to -10 \kms. The CO broad lines and H$_2$ emission show a spatial correlation, but in the local details, the strongest peaks may differ.}
\label{hb3coh2south}
\end{figure*}

%Figure10 
\begin{figure*}
\includegraphics[scale=1,width=17.5truecm,angle=0]{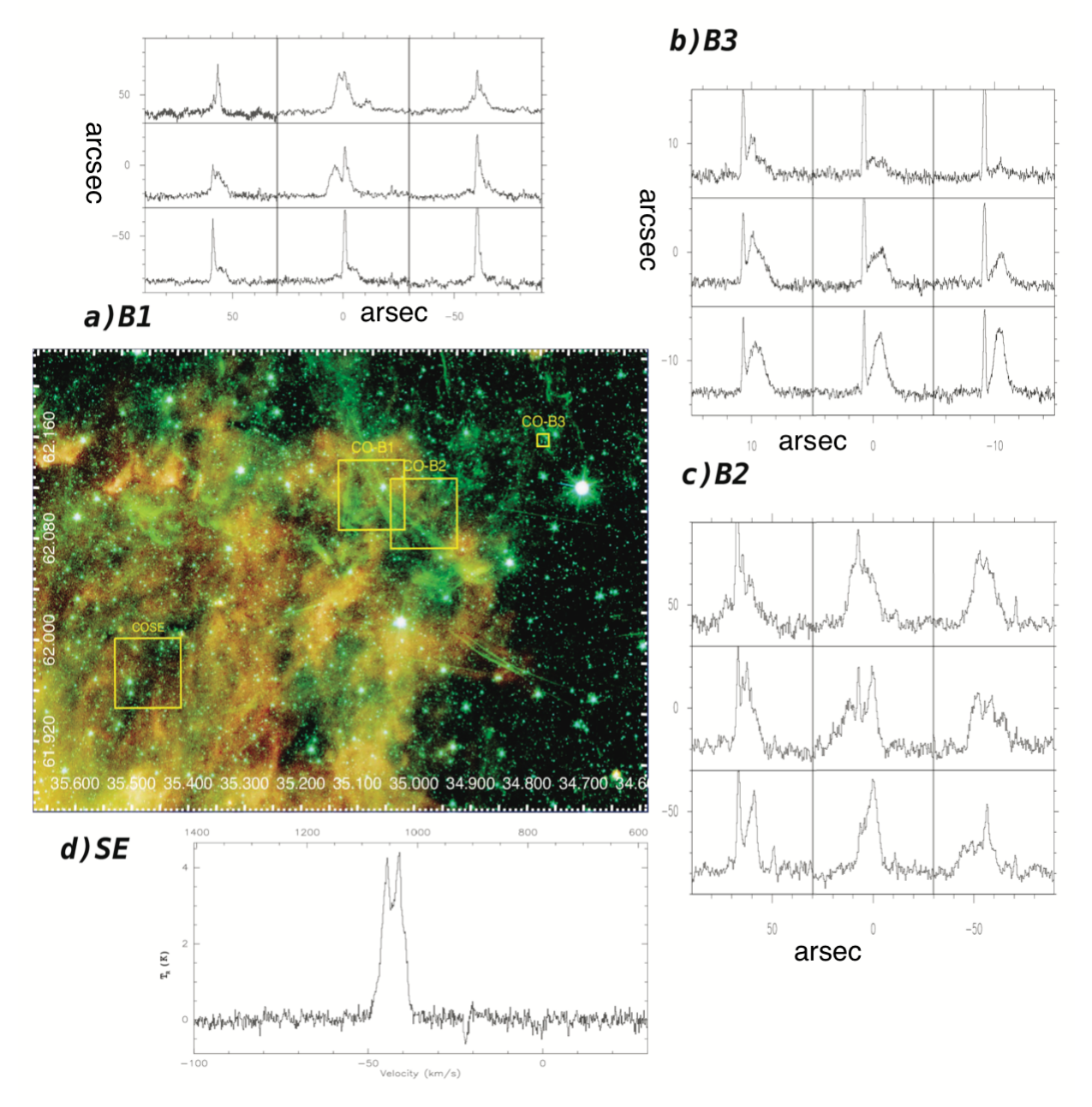}
\caption{Spatial distribution of molecular CO toward the southeast portion of HB\,3 and W3 complex (the exact locations are marked as yellow boxes on a large map of HB 3 in Figure \ref{hb3roadmap} and summarized in Table \ref{Tpos}). GRID spectra (in panels of (a, b, and c) which have the unit of arcsec in x- and y-axis like Figures \ref{hb3m3cospecnew}a and \ref{hb3m3cospecnew}b) of CO(3-2) are toward B1 $(a)$,  B2 $(c)$, and B3 $(b)$ region. %CO(2-1) spectrum is toward the SE (COSE) position in infrared dark clouds. 
The CO positions (as yellow boxes) are marked on the \spitzer\ and WISE three-color images (center).
Panels: 
$(a)$ CO(3-2) GRID spectra toward B1 region, with the X-axis velocity from -100 to +30 \kms\ 
$(b)$ CO(3-2) GRID spectra toward B3 region (the middle of the SNR), with a velocity from -70 to 0 \kms,
$(c)$ CO(3-2) GRID spectra toward the B2 region,  velocity ranging from -100 to +30 \kms.
$(d)$ A CO(2-1) spectrum toward SE position (see Table \ref{Tpos}) in velocity. The values on the upper side are detector channel numbers.}
\label{hb3coh2boundary}
\end{figure*}
\subsection{H$_2$ Line Contribution to Mid-IR images}
\label{Sh2model}

 We measured the fluxes at a few bright peaks in the {\it Spitzer} images (for 3$\times$3 pixels with a 1.2$''$pixel size) and averaged the fluxes in the WIRC H$_2$ images for the same area. We selected small areas which do not include any stars. Table \ref{Th2spitzerflux} summarizes the fluxes of the narrow-band H$_2$ (2.12$\mu$m) image and \spitzer\ images at 3.6 and 4.5$\mu$m. The filaments in WIRC H$_2$ image show that they are thinner than \spitzer\ resolution of 3$''$, but the measurements for smaller areas introduce large errors. The WIRC narrow-band 2.12$\mu$m fluxes (\ergsb) is converted to MJy\,sr$^{-1}$ after accounting for the filter transmission curve. The background emission at these short wavelengths is low and and it is near-zero measured from the lowest emission in the images. The archival \spitzer\ images are zodi-subtracted. The observed ratios of IRAC-2/H$_2$ (2.12$\mu$m) and IRAC-2/IRAC-1 are 0.8 - 1.2 and 0.9 - 1.9, respectively. The filament $A$ shows brighter than other $B - E$ filaments and shows higher ratios of IRAC-2/H$_2$ and IRAC-2/IRAC1 than others. The filament $A$ appears the combination of multiple filaments, as shown in the bright emission (Figure \ref{hb3wircb}d). Other filaments ($B - E$) are isolated thin filaments. \cite{reach06} show that the IRAC-2/IRAC-1 ratios of another 10 molecular interacting SNRs range from 1 to  3 \citep{reach06}.

\cite{reach06} suggest that the IRAC emission is largely H$_2$ lines in IRAC-1 and IRAC2 bands from H$_2$. IRAC-1 is composed of O(5) and S(13) H$_2$ lines, and the IRAC-2 is composed of S(11) and S(9) H$_2$ lines and possibly from CO emission. \citet{neufeld08} examine the IRAC maps of the supernova remnant IC 443 and suggest that the IRAC-1 and IRAC-2 are dominated by H$_2$ $v$ = $1-0~O(5)$ and $v$ = $0-0~S(9)$ based on modeling of the H$_2$ excitation, and CO vibrational emissions do not contribute significantly to the observed IRAC-2 intensity since CO requires very high densities ($>$ 10$^{7-8}$ cm$^{-3}$). AKARI IRC spectra cover the wavelengths of IRAC-1 and IRAC-2 bands, and \cite{shinn11} detected H$_2$ lines from IC\,443. The IRC-detected H$_2$ lines within the IRAC-1 band are  $1-0~O(5)$ at 3.23\mic, 0-0 S(15) at 3.62\mic, 1-0 O(7) at 3.81\mic, and 0-0 S(13) at 3.84\mic, and the lines in the IRAC-2 are 0-0 S(11) at 4.18\mic\, 0-0 S(10) at 4.41\mic\, 1-0 O(9) at 4.58\mic, and 0-0 S(9) at 4.69\mic. \cite{neufeld08} suggest that the relative strengths of IRAC band emission are consistent with pure H$_2$ emission from shocked material with a power-law distribution of gas temperatures. The IRAC-2/IRAC-1 ratios depend on the density  \citep{neufeld08, shinn11, shinn12} as well as the slope of the temperature distribution.

In summary, \spitzer\ IRAC images at 3.6 and 4.5$\mu$m can be explained by bright lines of molecular hydrogen for the shocked filaments and would not require a contribution from the continuum. Future sensitive spectroscopic follow-up observations for IRAC-1 and IRAC-2 (from 3.2\mic\ to 5\mic), such as using JWST, may be able to detect faint H$_2$ lines and test how the parameters of H$_2$ excitation diagram in HB\,3, densities, or temperature distribution are different from those of IC 443, which can lead to an understanding of the shock properties and astrochemistry to form water.

%Figure11 
\begin{figure}
\includegraphics[scale=0.72,width=5.truecm,angle=270]{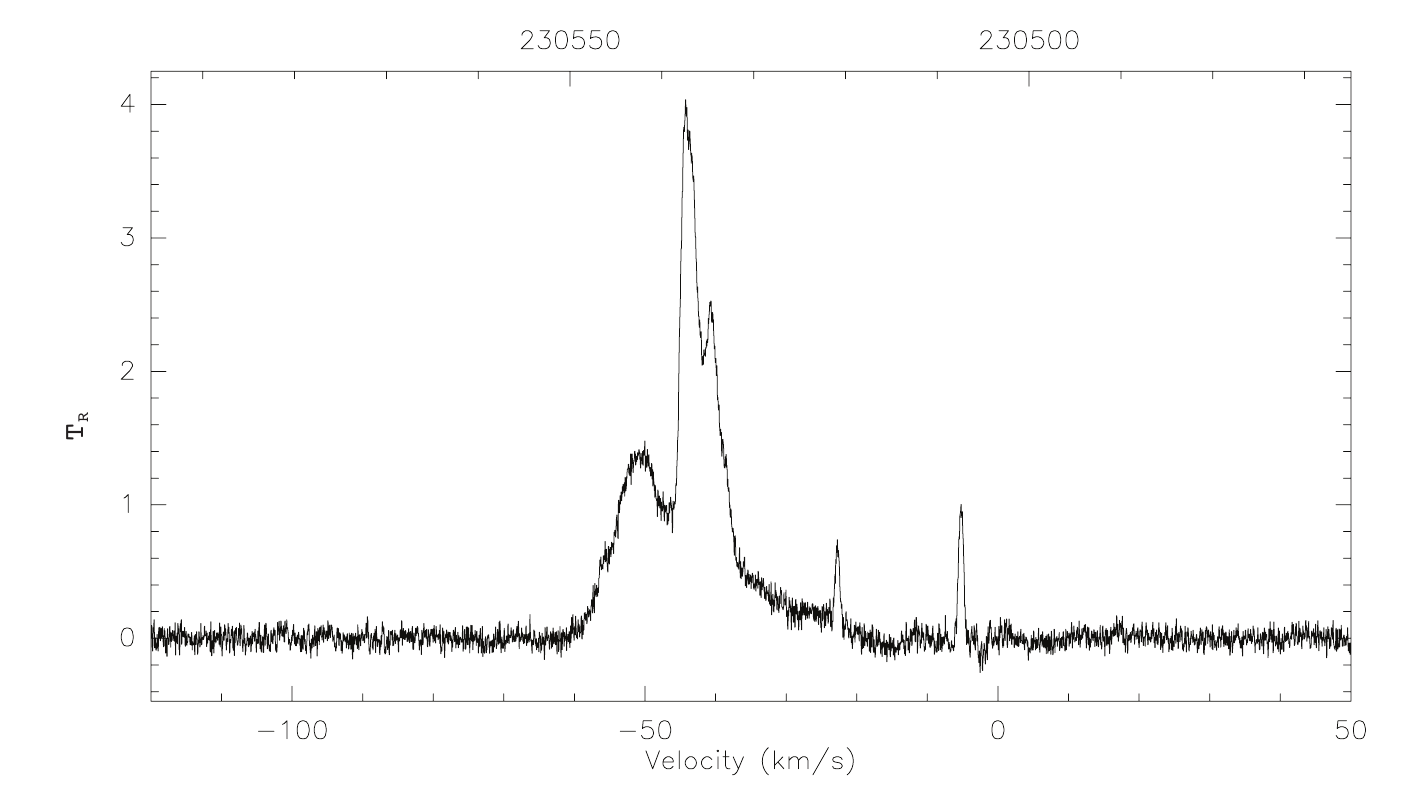}
\caption{12-Meter CO(1-0) spectrum of HB\,3 B1 (CO(3-2) spectrum is shown in Figure \ref{hb3coh2boundary}a) with an offset of (+10,+20) shows a broad CO line with a FWHM of 18 \kms\ with high signal-to-noise.
The numbers shown on the upper side are frequencies in MHz.}
\label{hb3m3cospecnew}
\end{figure}

%Figure12 
\begin{figure}
\includegraphics[scale=1,width=7.truecm,angle=0]{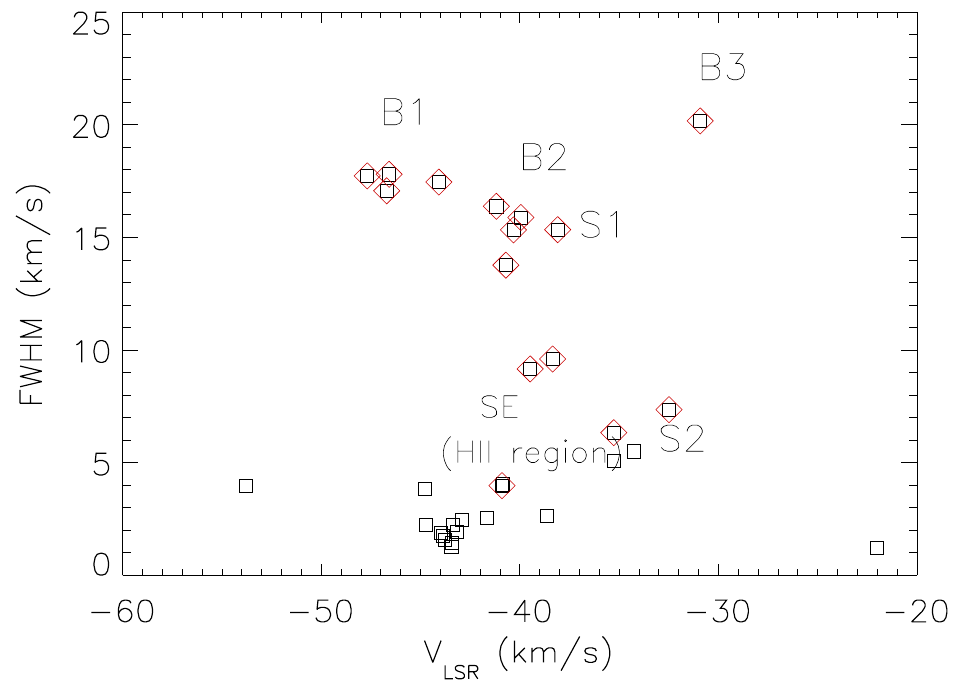}
\caption{Distribution of LSR velocity and the Full-Width-Half-Maximum (FWHM) at different positions of HB\,3, including the FWHMs from representative spectra shown in Figures \ref{hb3coh2south} and \ref{hb3coh2boundary} in addition to those in Table \ref{TCOlines}. The first component (in red) are broad lines except in the SE (HII) region.}
\label{hb3velfwhm}
\end{figure}

\begin{table*}
\caption[]{Summary of CO line Properties \label{TCOlines}}
\begin{center}
\begin{tabular}{llrlllrrlll}
\hline \hline
Position &  line & Frequency & offset  & compo-& V$_{lsr}$ &   FWHM & $\int Tdv$ &RMS & t$_{int}$ & Figs. \\
         &             &     &  ($''$,$''$)   &nent     &    & (km s$^{-1}$) & (K km s$^{-1}$) &(K) & (min) & \\
\hline
S2 & CO(3-2) & 345.7959 & (0,0)&1 &  -32.48$\pm$0.16 & 7.35$\pm$0.41 & 11.16$\pm$0.48 &0.17 &5 & \ref{hb3allcospec}, \ref{hb3coh2south} \\
           & &&                &2 & -41.67$\pm$0.04 & 2.55$\pm$0.14 & 7.26$\pm$0.31 & & \\
S1 & CO(3-2) & 345.7959 & (0,-60)  &1 & -38.09$\pm$0.26 & 15.34$\pm$0.61 & 24.32$\pm$0.83 & 0.14 & 5  & \ref{hb3coh2south}\\
   &         &          &          &2 & -53.80$\pm$0.22 & 3.98$\pm$0.56  & 3.63$\pm$0.46 &      &     &  \\
\hline
B3  & CO(3-2) & 345.7959   & (10,-10) &1& -30.91$\pm$0.15 & 20.17$\pm$0.36 & 47.76$\pm$0.60 & 0.12  & 6 & \ref{hb3allcospec}, \ref{hb3coh2boundary} \\
                &&   & &2 & -44.73$\pm$0.03 & 2.25$\pm$0.08  & 6.79$\pm$0.23 & \\
B2  & CO(3-2) & 345.7959& (0,+60)&1 & -41.18$\pm$0.15 & 16.38$\pm$0.35& 43.74$\pm$0.78& 0.18  & 5 & \ref{hb3allcospec}, \ref{hb3coh2boundary} \\
               & &&        & 2 &-43.89$\pm$0.08 & 1.73$\pm$0.05 &2.75$\pm$0.25 &&\\
B1   & CO(3-2) & 345.7959 & (0,-60) &1 & -39.94$\pm$0.37 & 15.88$\pm$0.82 &19.87$\pm$0.95& 0.21& 6 & \ref{hb3coh2boundary}\\
                  &   &&   &2   &-43.36$\pm$0.03   & 2.24$\pm$0.06  & 13.35$\pm$0.41 &  &  \\
B1  & CO(3-2) & 345.7959 & (0,0) & 1 &-47.68$\pm$0.16  & 17.73$\pm$0.27 & 43.08$\pm$0.70 &0.16 & 5 \\
    &         &          &       & 2 &-43.17$\pm$0.06  & 1.92$\pm$0.13  & 4.94$\pm$0.31 \\
B1 & CO(2-1) & 230.5379& (+10,+20) &1 & -46.58$\pm$0.03 & 17.80$\pm$0.03 & 23.18$\pm$0.02 &  0.06 & 53 &\ref{hb3m3cospecnew}  \\
   &&&             &2 &          -43.96$\pm$0.03 & 1.87$\pm$0.03 & 5.32$\pm$0.02 &       &  \\
   &&&             &3 &       -40.86$\pm$0.03& 4.03$\pm$0.03 & 6.05$\pm$0.02 &    &  \\
SE   &  CO(2-1) & 230.5379 & (0,0)&1 & -40.89$\pm$0.03& 3.97$\pm$0.28& 13.74$\pm$0.90 & 0.16& 1 & \ref{hb3coh2boundary} \\
     & &&      &2&-44.76$\pm$0.06 & 3.82$\pm$0.37 & 11.67$\pm$0.94 &  & \\
\hline \hline
\end{tabular}
\end{center}
\renewcommand{\baselinestretch}{0.9}
\end{table*}

%Figure13
\begin{figure*}
\begin{center}\includegraphics[width=15truecm]{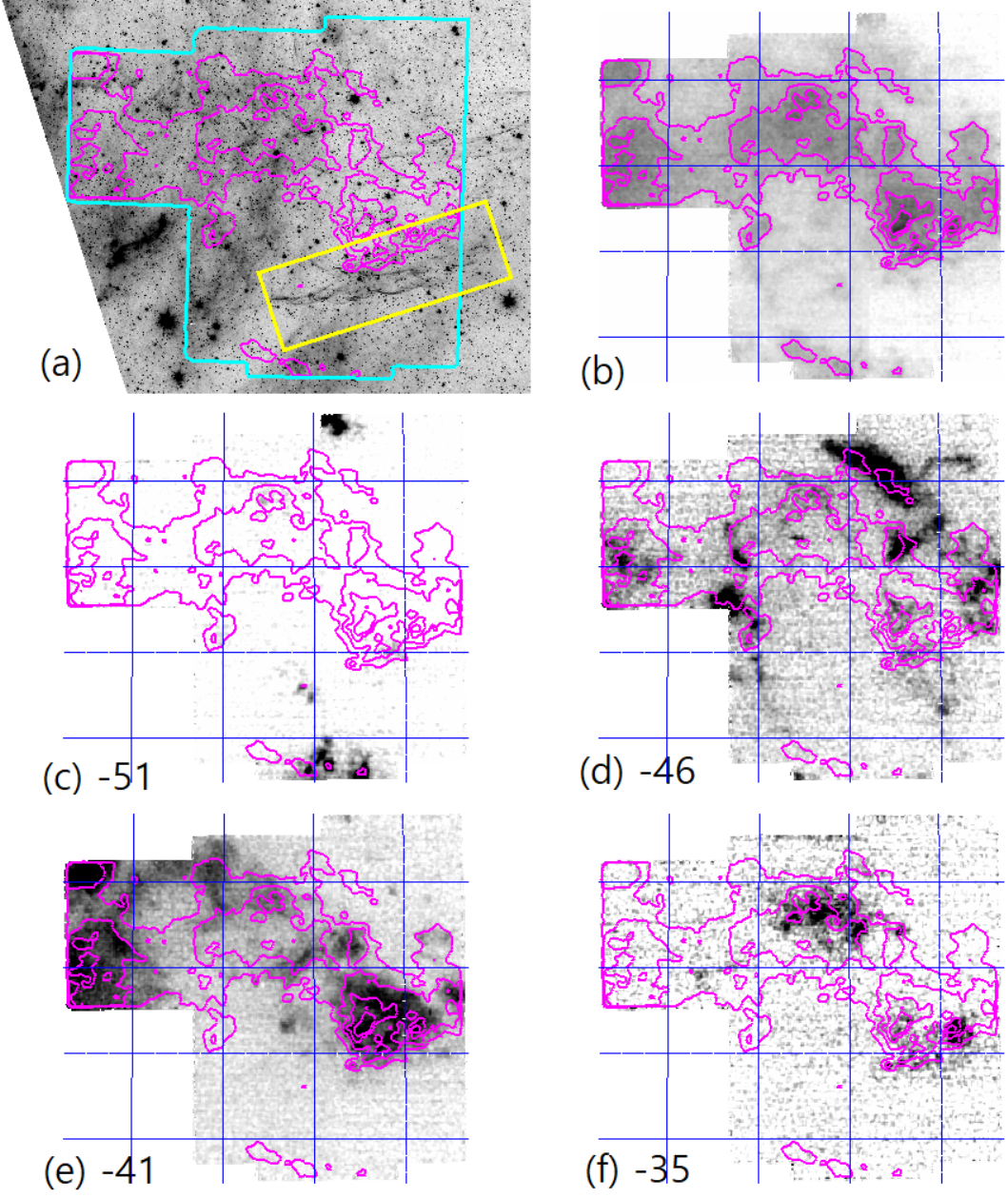}
\end{center}
\caption{Kitt Peak 12-Meter CO(2-1) channel maps together with the \spitzer\ 4.5 $\mu$m mosaic toward the south and south-eastern regions of HB\,3; $(a)$ \spitzer\ 4.5 \mic\ superposed on molecular CO integrated intensity contours over V$_{LST}$ = -53 to -35 km/s. $(b)$ CO integrated map (from -53 to -35 \kms), 
$(c)-(f)$ = channel maps ($\Delta~v$ = 0.65 \kms) centered at -51, -46, -41, and -35 \kms, respectively. The contours are those of integrated intensities at 20, 30, 40, 50 K \kms. The yellow box in (a) is mentioned in the text. The -35 \kms\ CO clouds coincide with the observed H$_2$ emission at the southern (outer) shell. 
In frames (b)--(f), coordinate grids are shown at every $1^{\rm m}$ $30^{\rm s}$ in R.A. from $2^{\rm h}$ $18^{\rm m}$ to $2^{\rm h}$ $22^{\rm m}$ $30{\rm s}$ and
at every $10'$ in Dec. from $+61^\circ$ $40'$ to $+62^\circ$ $10'$. The CO image is centered on R.A.\ $2^{\rm h} 20^{\rm m} 17^{\rm s}$ and Dec.\ $+61^\circ$55$^{\prime}37.5^{\prime \prime}$ (J2000) with a FOV of 46.8$'$x42.8$'$.}
\label{hb3SCOmap}
\end{figure*}

\section{Millimeter and Submillimeter CO Emission}

\subsection{Broad CO spectra}

Figure \ref{hb3roadmap} shows six regions that we observed CO(3-2), CO (2-1) and $^{13}$CO(2-1) lines using HHSMT. These positions were selected based on radio and X-ray images before the \spitzer\ and WISE images were available. The southern (S) positions are along the southern shell and the boundary (B) positions are close to the boundaries between the SNR HB\,3 in the south-eastern direction and the HII regions of W3 complex. The regions of S1, S2, B1, B2, and B3 in Figure \ref{hb3roadmap}, show broad molecular lines in CO(2-1). The representative, resolved spectra are shown in Figure \ref{hb3allcospec}, and the grid maps are shown in Figures \ref{hb3coh2south} and \ref{hb3coh2boundary}. A spectrum with the broadest line from the GRID map of each region was fit using two component gaussian profiles with wide and narrow lines; the fit results are summarized in Table \ref{TCOlines}. 

When we compare the CO GRID map and H$_2$ emission as shown in Figures \ref{hb3coh2south} and \ref{hb3coh2boundary}, it is clear the interaction between HB\,3 and dense molecular clouds based on the detections of broad CO(3-2), CO (2-1) and $^{13}$CO(2-1) molecular lines from a few positions (S1, S2, B1, B2 and B3) of HB\,3 along the H$_2$ emitting region. The Kitt Peak CO(1-0) line toward position B1 is shown in Figure \ref{hb3m3cospecnew}. The broad lines show line widths of 7-21 \kms.  The detection of such broad lines is unambiguous, dynamic evidence showing that the SNR HB\,3 is interacting with dense gas. Such broad CO lines require the FWHM to be greater than 6 \kms\ \citep{kilpatrick16} for kinematic (shock) broadening. The CO broad lines are detected from  two branches of the southern shell where H$_2$ emission is detected and prominent. Whereas in contrast, CO emission toward the HII region of W3 complex detects only a width of 3 \kms\, which indicates that the SN shock did not propagate into the eastern HII region.

Figure \ref{hb3velfwhm} shows the kinematic distribution between the LSR velocity and the FWHM width shown in Table \ref{TCOlines}. The W3 HII regions have a systematic velocity of -41 \kms. The narrow components of both HII regions and the SNR HB\,3 are at a velocity of -41 to -46 \kms. The broad lines along the southern shell of HB\,3 appear at velocities of -32 to -39 \kms, and the boundary regions between HB\,3 and the HII region are at velocities of -40 and -47 \kms, which probably include the material moving toward us and receding from us. The CO clouds toward the position of B3 are at a velocity of -31 \kms which is gas that is mainly receding from us (it is redshifted relative to -45 \kms).

\subsection{CO broad lines and CO maps in the southern shell}

The CO broad lines are located at the lower part of the S1 region where there is brighter H$_2$ emission as well, similar to the central regions. An isolated, thin H$_2$ filament in the northern part of S2 shows CO broad lines with a relatively small, FWHM 7.4 \kms, which may imply that this isolated shock is spatially resolved. The broader lines of 15-20 \kms\ may be due to a shock velocity of the same order, but alternatively for some cases, the lines could be multiple shocks of smaller broad lines, closer to $\sim$8 \kms. Further follow-up observations with a higher spatial resolution may be needed to distinguish between the two components. \citet{zhou16} have presented broad CO lines from the southeastern part of HB\,3 and suggested shock-cloud interaction in HB\,3. Since the CO line profiles \citep{zhou16} differ from the ones presented in this paper, they may cover somewhat different area, but their position C seems to be from the eastern shell, where there is bright H$_2$ emission. One-to-one correspondence between shocked H$_2$ emission and broad CO lines are suggested from faint H$_2$ emission in IC 443 \citep{rho01}. Thus we can expect this correlation for the shocked H$_2$ gas, which appear along with the radio shell of HB\,3.

\subsection{Large-scale CO maps and comparison with H$_2$ gas}
Figure \ref{hb3SCOmap} shows CO(2-1) maps with velocity channel maps, compared with the \spitzer\ H$_2$ map. The upper H$_2$ filament (in the yellow box of Figure \ref{hb3SCOmap} marked as ``southern outer-shell" in Figure \ref{hb3roadmap}) matches very well to the -35 \kms\ (or -41 \kms) cloud. It indicates the H$_2$ emission is shocked gas due to the interaction of the SNR HB\,3 and cold molecular clouds. The velocity at -35 \kms\ is the same velocity range observed toward S1 and S2 clouds in the CO(3-2) HHSMT spectra as shown in Figures \ref{hb3velfwhm} and \ref{hb3coh2south}. \cite{zhou16} concluded that the remnant is located at the nearside of of the molecular cloud because redshifted CO wings are observed. We find the CO clouds of the southern shell (S1 and S2) and central regions (B3) are at velocities -30 to -36 \kms and the velocity channel map at -35 \kms\ also infers a similar range. The shocks in Regions B1-B3 and Regions S1-S2 are progressing in somewhat different directions, generally on the front and far side of the SNR.

%Figure14
\begin{figure}
\includegraphics[height=5.5truecm,width=7.truecm]{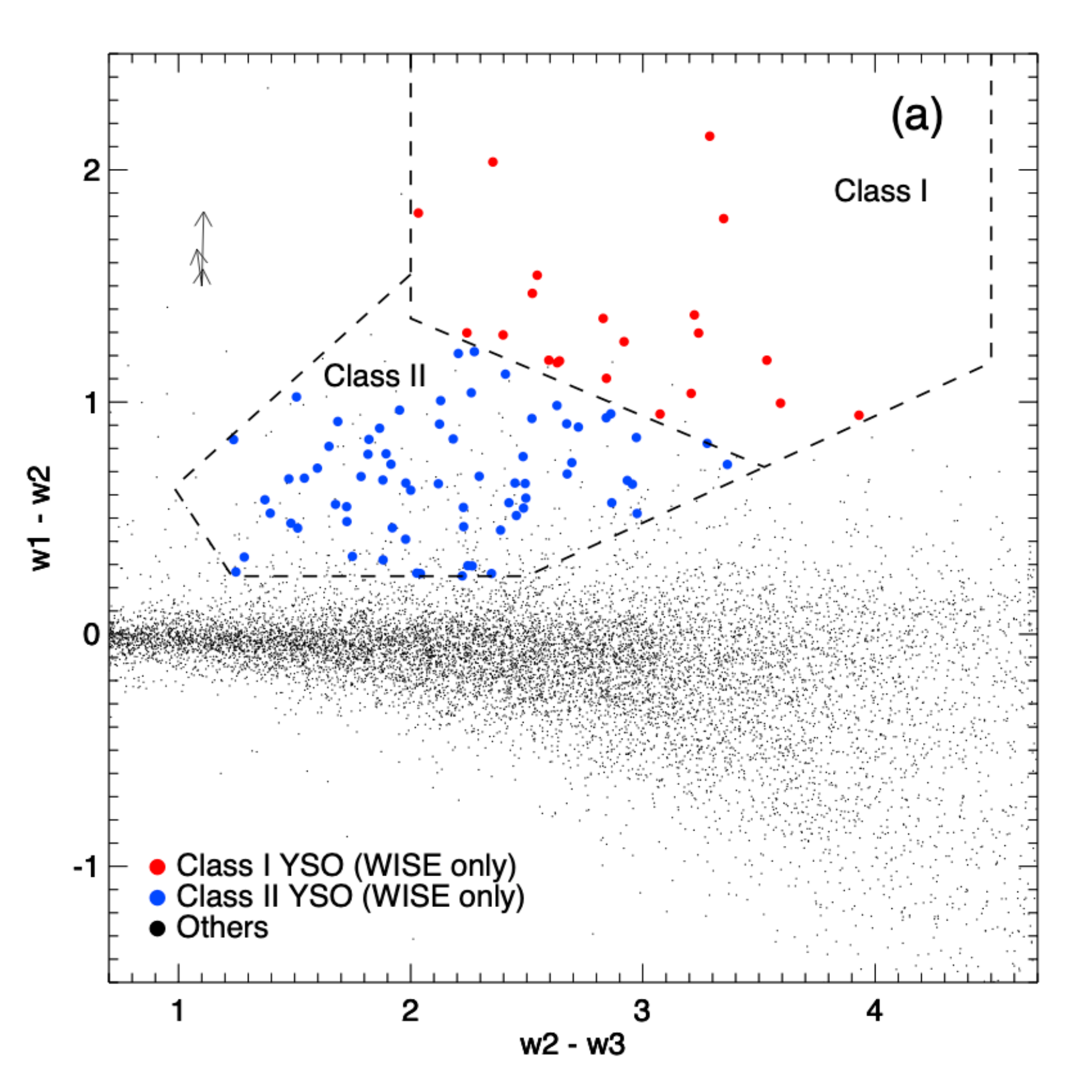}
\includegraphics[height=5.5truecm,width=7.truecm]{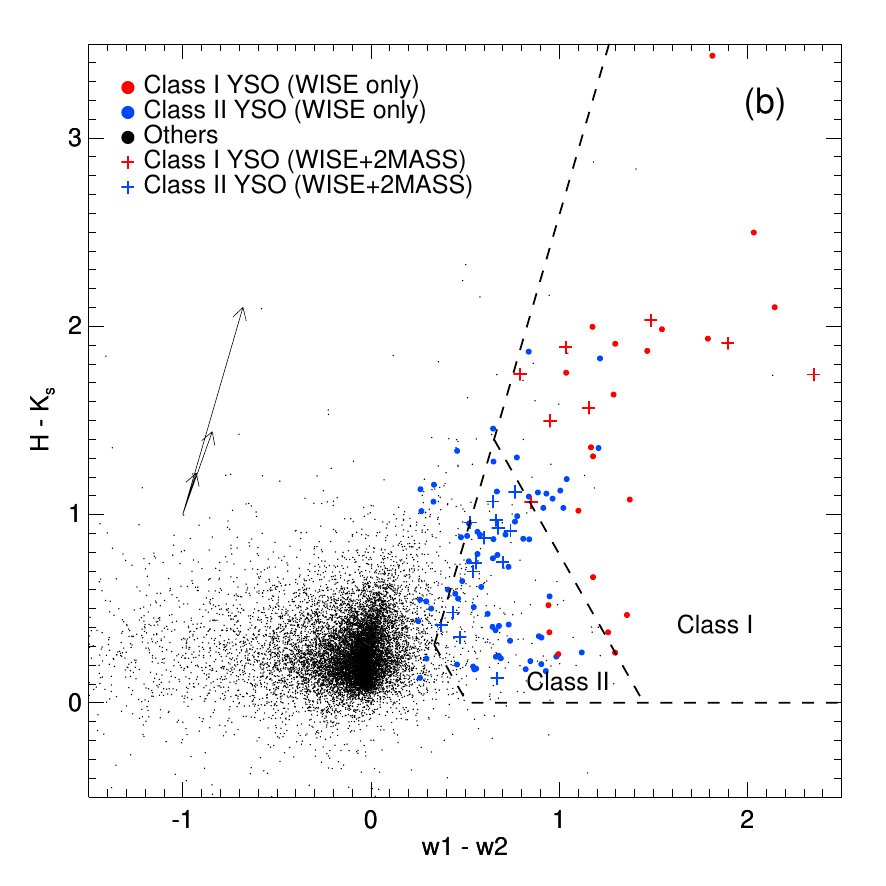}
\caption{{\it (a)} The mid-IR color-color diagram of HB\,3 region determining class~I and class~II YSOs based on the criteria (marked in dotted boxes) of \citet{koenig14} using the WISE photometry only. The red, blue, and black dots show class~I YSO, class~II YSO, and other sources, respectively. The black data points within the defined boxes are background star-forming galaxies or Galactic AGB stars \citep{jarrett19}. {\it (b)} YSOs in the region viewed in a color-color diagram of 2MASS-WISE photometry data. The red and blue crosses indicate the class~I and class~II YSOs defined from the 2MASS-WISE color-color diagram, while the colored dots are from WISE color-color diagram. The dashed polygons of both plots are the color-color diagram criteria for class~I and class~II YSOs. Arrows in both plots are extinction vectors of A$_{Ks}$ = 0.4, 0.8 and 2. 
}
\label{fig:ccd}
\end{figure}

\subsection{CO line profiles at the boundary between HB\,3 and the W3 complex}

Figure \ref{hb3coh2boundary} shows a CO(2-1) spectrum toward the SE direction of HB\,3 where it is in the HII region;  the area is marked in Figure \ref{hb3roadmap}. The CO line shows two narrow components (see Table \ref{TCOlines}). We covered  a large area of 3.5$'$ $\times$3.5$'$ toward this direction, and no broad CO line is detected. From the mid-IR WISE and GLIMPSE maps, the position is clearly located in the nearby W3 complex. Hence, the CO spectra are indicative of the H~II region. The line widths are about 3-4 \kms. WISE and GLIMPSE imaging do not show any shocked molecular hydrogen emission from this area, indicating that the SNR boundary is further north-west of the HII region. H$_2$ emitting regions follow the boundary of the SNR HB\,3. We also observed an eastern position (EI; the inner part of the eastern shell), but inside the bright eastern H$_2$ filaments, as shown in Figure \ref{hb3roadmap} that covers 8$'$ $\times$5$'$, and no CO line is detected toward this position. Moreover, this area coincides with no H$_2$ emission in both \spitzer\ IRAC-2 and WISE-$w2$ images; see Figures \ref{hb3spitzeriracall} or \ref{hb3wisezoom}.

The spacing of our CO GRID spectra  toward the B3 region in Figure \ref{hb3coh2boundary} is 10$''$, and the spectra show the broadest CO lines. Their H$_2$ emission morphology is knotty and clumpy as shown in Figures \ref{hb3spitzeriracall} and \ref{hb3roadmap}. They are either gas that is moving toward and away from us (i.e. less material in tangential) which causes broader CO lines; or more likely, the emission is due to C-shock origin with a higher pre-shock density. The B1 region has both filamentary and knotty H$_2$ emission as shown in Figure \ref{hb3coh2boundary}.

The B2 region exhibits the most complicated CO line profiles, as shown in Figure \ref{hb3coh2boundary}. The velocities vary considerably within the 180$''$x180$''$ area. It is likely the clouds are interacting with both the SNR and the HII region. In contrast, the velocities of the line profiles of the B3 region remain the same, but the strengths of the WISE mid-IR components significantly vary across the area.

\section{Star-forming Activity and Young Stellar Objects}

We identify star-forming activities in and near HB~3 and W3 complex by investigating the distribution of Class~I and Class~II young-stellar objects (YSOs) using two color-color diagrams from the {\it 2MASS} \citep{skrutskie06} and {\it WISE} point source catalogs. We adopt the YSO criteria described in \citet[][KL14, hereafter]{koenig14} as in \cite{zhou16}. The source classifying algorithm (KL14) first defines the star-forming galaxies \citep[SFGs;][]{jarrett19} and asymptotic giant branches (AGBs) under their specific criteria in color-color diagrams (equations 5-8 and 9-11 in KL14 for SFGs and AGBs, respectively). They then select Class~I and II YSOs with WISE color-color diagram. The criteria for Class~I YSOs with WISE data only are $w2-w3 > 2$ and $w1-w2 > -0.42 \times (w2-w3)+2.2$ and $w1-w2 > 0.46 \times (w2-w3)-0.9$ and $w2-w3 < 4.5$. For Class II YSOs,  the criteria are the following: $w1-w2 > 0.25$ and $w1-w2 < 0.9 \times (w2-w3) - 0.25$ and $w1-w2 > -1.5 \times (w2-w3) + 2.1$ and $w1-w2 > 0.46 \times (w2-w3) -0.9$ and $w2-w3 < 4.5$. The locations of Class~I and II YSOs are marked on the WISE color-color diagram in Figure~\ref{fig:ccd}a. The remaining sources are investigated again with the photometry measurement of 2MASS and WISE so that the criteria of Class~II YSOs are $H - K_S > 0.0$, $H - K_S > -1.76 \times (w1 - w2) + 0.9$, $H - K_S < (0.55/0.16) \times (w1 - w2) - 0.85$, and $w1 \le 13.0$. In order to determine Class~I YSOs, we add a criterion $H - K_S > -1.76 \times (w1 - w2) + 2.55$. The locations of Class~I and II YSOs are marked on the {\it 2MASS} and {\it WISE} color-color diagram in Figure~\ref{fig:ccd}b.

\begin{figure*}
\begin{center}
\includegraphics[scale=0.7,width=14.5truecm,angle=0]{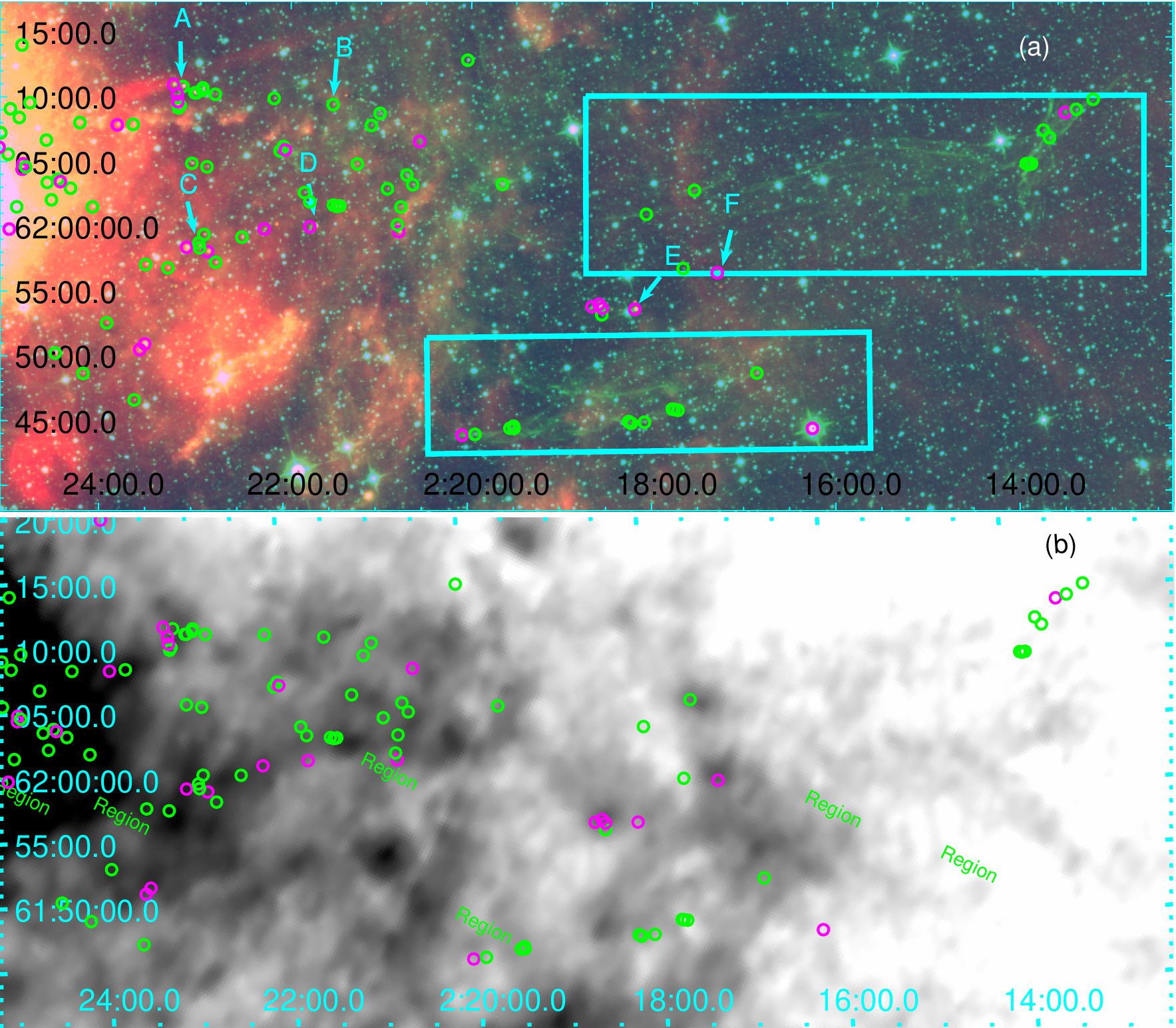}
\end{center}
\caption{{\it (a)} WISE three-color image ($w1$ in blue, $w2$ in green and $w4$ in red) highlighting star-formation sites
(marked as A-F) toward the boundary of the SE of HB\,3 and the H~II region
W3 and the southern H$_2$ filament. The star-forming regions marked as  A
and B are located at the boundaries between the SNR and the W3 complex, C
and D are at the edge/in the HII regions, and E and F are outside the W3
complex, which may be in the dark clouds and associated with SNR feedback.
The red and green circles indicate the positions of
class~I and class~II YSOs, respectively. 
Class ~I YSOs tend to locate
in dark clouds where they are born. The southern inner (the upper box) and outer (the lower
box) shell regions are marked as a box (see also Figure \ref{hb3roadmap}.)
{\it (b)} YSOs of class~I  (in magenta) and class~II (in green) are marked on the
AKARI 160\mic\ image. Class~I YSOs are located along the long filaments or
clumps of bright emission of 160\mic\ map.}
\label{hb3sf}
\end{figure*}

Figure \ref{hb3sf}a shows SF activities (marked as Regions A-F) around the SNR HB\,3. Figure~\ref{hb3sf}a shows the locations of all Class~I and II YSOs in the inspected field on the WISE three-color map.  The distribution of YSOs shows that some of YSOs are in the infrared dark clouds. There are three types of SF activities; i) in the dark clouds near W3 complex and SNR with the examples of Regions C and D; ii) at the boundary between HB\,3 and W3 H\,II region complex along the $``$W" shape of H$_2$ emitting area, with the examples of Regions A and B; and iii) at the H$_2$ emitting arc structures of SN southern shell showed by YSOs between the southern inner and outer shell (see Figure~\ref{hb3roadmap}), with examples of Regions $E$ and $F$. 
Figure \ref{hb3sf}b shows the locations of YSOs marked on the AKARI 160\mic\ image. Class~I YSOs are located along the long filaments or clumps of bright emission in the 160\mic\ map, while Class~II YSOs are more scattered relative to the 160\mic\ emission.

The southern shell structures in the FOV of Figure \ref{hb3sf} (marked as two boxes) show that the blast-wave propagates from north to south (the center of SNR is in the north/northeast of the shell structure, as shown in the large FOV of Figure \ref{hb3roadmap}) and more Class~II populations are present along with the shell structures. The shock wave of HB\,3 may have plowed through YSOs and located them at the southern shell of the SNR. Recent studies show that the relatively older YSO populations have a higher spatial difference to the birthplace \citep{elmegreen18} or the higher kinematic difference from the dense molecular clouds \citep{lim21}. The HB\,3 regions also have been assumed to experience the repeated powerful feedback from the massive progenitor of HB\,3 \citep[e.g.,][]{zhou16}. Such events could result in a more reduced spatial and kinematic correlation between the Class~II YSOs and mother clouds. Moreover, since Class~I YSOs are associated with dense molecular clouds, it would be harder for the blast wave to plow the dense gas, and instead, the shock wave would slow down where we observe both Class~I and II YSOs at the boundary regions between HB\,3 and W3 Complex.
One needs to bear in mind that the comparison of separate evolutionary tracers of such SF activities, e.g., virial parameter ($\alpha_{\rm vir}$) vs. luminosity - mass ratio $(L/M)$ \citep{lim19} or unambiguous identification of YSOs with follow-up spectroscopy is still needed  to support our conclusion.

%Figure15 

\section{Physical conditions of shocked gas in HB~3}

\subsection{Resolved Shock Structures in H$_2$ and CO}

The H$_2$ emission of HB\,3 exhibits largely thin filamentary structures and knotty structures. The thickness is 1 or 2 WIRC pixels (one pixel size is 0.25$''$), which is comparable to 0.8-1.6$\times$10$^{16}$ cm for the distance of HB\,3. The filamentary H$_2$ emission is unlikely to arise from very dense cores (n$_0$ $>$10$^{5}$ cm$^{-3}$), because the filaments are coherent and long ($>$ a few pc), as discussed in \citet{reach05}. The filamentary structures require that the pre-shock medium be very uniform and probably relatively low ($\sim$10$^{3-4}$ cm$^{-3}$; see the application of shock models in next Section) gas density. We compare H$_2$ emission with shock models. The H$_2$ brightness of 0.6-2$\times$10$^{-5}$ \ergsb\ is consistent with a J-shock with a low density of  $n_0$ = 10$^3$ cm$^{-3}$ \citep{hollenbach89}. Knotty structures are more than a factor of 3 brighter than those of filamentary H$_2$ emission; see  Figure \ref{hb3wircb}d. It is likely that the shocks are moving along the line of sight, or they are from multiple shocks including the shocks moving in the light of sight.

The primary narrow component is at a velocity of  -42 \kms\, and is likely pre-shocked gas. The V$_{lsr}$ of broad CO lines range from -30 and -46 \kms, which is a large velocity difference, as shown in Table \ref{TCOlines}. In the GRID maps, we see the broad line structures change within 3$\times$3 maps.

The CO observations of HB\,3 is previous presented by \citet{kilpatrick16}, the authors observed  the broad molecular line region toward HB\,3 and the W3(OH) HII region. \citet{kilpatrick16} concluded that the broad lines are not associated with the SNR.   Our first CO data were observed before the infrared H$_2$ images were available. However, when the mid-infrared GLIMPSE and WISE data are factored into this analysis, it is then clear that the H$_2$ emission reveals the regions where the SNR HB\,3 is clearly interacting with the molecular clouds, and the regions where the SNR HB\,3 is interacting with the HII region of W3. The broad CO lines are associated with the H$_2$ radiative cooling lines at the shock front, as shown in Figures \ref{hb3coh2south} - \ref{hb3m3cospecnew}. Our observations show a strong correlation between H$_2$ and broad CO lines, not unlike the case of the SNR IC\,443 where one-to-one correspondence between H$_2$ and broad CO are reported \citep{rho01}.  The shock-excited molecular hydrogen traces where the gas has been heated, kinematically broadening the observed CO emission lines. 
The fact that the SNR HB\,3 is interacting with nearby molecular clouds suggests that the progenitor of HB\,3 is born or associated with a GMC.
However, the progenitor must be later than B0 \citep{rho94} in order to have direct interaction between the SNR and local molecular clouds, because more massive stars would have strong winds, of which wind-driven bubbles clear away the clouds.

The SNR HB\,3 shows very unique and interesting CO structures. First, the velocity of LSR significantly varies  across the field, as shown in Table \ref{TCOlines}, and Figure \ref{hb3velfwhm}. Second, the broad lines appear and disappear on small physical scales: Figure \ref{hb3coh2boundary} shows at 10 arcsec scales, the broad lines appear and disappear. Third, the H$_2$ imaging indicates that individual shocks are resolved, where the FWHM of CO line is small,  7 \kms\ at a few positions in HB\,3.

\subsection{Applications of Shock Models to HB\,3 Spectra}
\label{Sshockmodel}

In this section, we model the line profile of the CO(3-2) transition and interpret to what is observed. The intensity of the CO(3-2) transition along the line-of-sight $s$ is obtained by solving the radiative transfer

\begin{equation}
    \frac{dI_{\nu}(s,\theta)}{ds} = \epsilon_{\nu}(s, \theta) -\kappa_{\nu}(s, \theta) I_{\nu}(s, \theta) 
\end{equation}
where $\epsilon$ and $\kappa$ are respectively the emissivity and absorption coefficient, and defined as
\begin{equation}
    \epsilon_{\nu}(s,\theta) = \frac{h\nu_{32}}{4\pi}n_{3}(s)A_{32}\phi_{\nu}(T_{\rm gas}(s), \theta)
\end{equation}
\begin{equation}
    \kappa_{\nu}(s,\theta) = \frac{h\nu_{32}}{4\pi}[n_{2}(s)B_{23}-n_{3}(s)B_{32}]\phi_{\nu}(T_{\rm gas}(s),\theta)
\end{equation}
where $\nu_{32}$ is the frequency of the CO(3-2) transition. $n_{2}$ and $n_{3}$ are the populations of the lower ($J=2$) and upper ($J=3$) excited stages of this transition. $A_{32}, B_{32}$ and $B_{23}$ are the Einstein coefficients of the spontaneous, stimulated emissions and absorption, respectively. $\phi_{\nu}(\theta)$ is thermal broadening of the line, which is a function of the viewing angle $\theta$ (i.e., the angle between the shock velocity and the line-of-sight). To obtain the line profile, we transform the argument $\nu$ into the velocity $v_{r}$ ($\phi_{\nu} \sim \phi(v_{r})$), which is
\begin{equation}
    \phi(v_{r}) = \frac{\lambda_{32}}{\sqrt{2\pi}\sigma}\exp{\left[-\frac{(v_{\rm obs}\cos \theta - v_{r})^{2}}{2\sigma^{2}}\right]}
\end{equation}
where $\lambda_{32}$ is the wavelength of the CO(3-2) transition, $\sigma \sim T^{0.5}_{\rm gas}$ is the gas thermal dispersion velocity, and $v_{\rm obs} = v_{\rm gas}(z)-v_{\rm shock} -v_{\rm preshock}$ is the shock velocity in the observer frame with $v_{\rm gas}(z)$ the gas velocity at postion $z$ in shock frame, $v_{\rm shock}$ the shock velocity, and $v_{\rm preshock}$ the preshock velocity.
The velocity ($v_{\rm gas}(z)$) and temperature ($T_{\rm gas}(z)$) of the shocked gas in the shock frame 
are calculated in a 1D plane-parallel Paris-Durham shock model (\citealt{lesaffre13}; \citealt{flower15}; \citealt{Gordard19}; \citealt{lehmann20}). We use non-stationary C-shocks that have dynamical ages shorter than the time required for a C-shock to reach the steady state, which is composed of a C-shock and a J-shock tail (referred as CJ-shocks). 
Note that in the case of a plane-parallel approximation, $s=z/\cos(\theta)$, with $z$ post shock position. Then, the emissivity ($\epsilon$) and the absorption coefficients ($\kappa$) are estimated by the large velocity gradient (LVG) assumption (\citealt{sobolev60}; \citealt{surdej78}).

%Figure16
\begin{figure}
    \centering
    \includegraphics[width=0.45\textwidth]{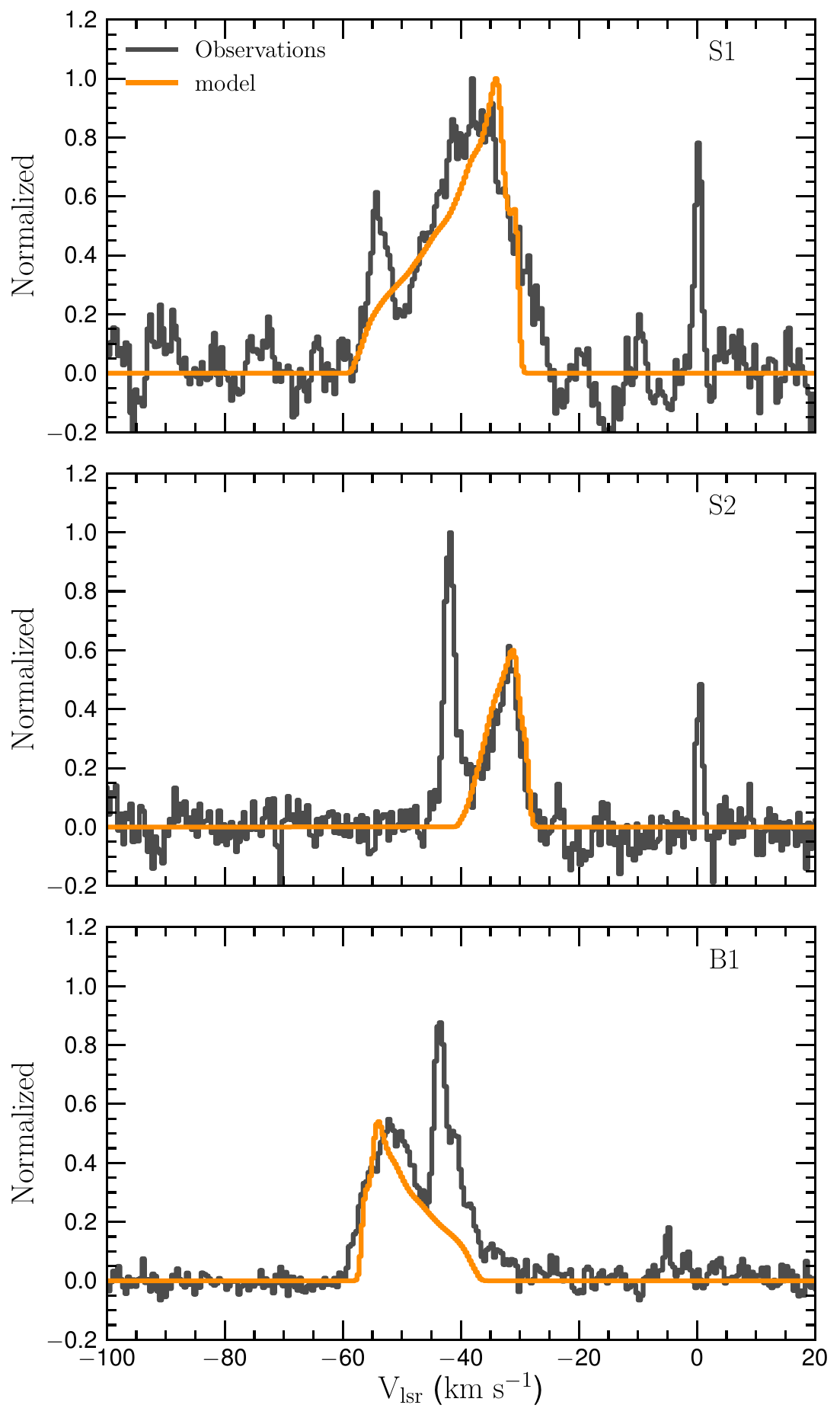}
    \caption{The best CO(3-2) line profile predicted from our shock model, compared to observations for the S1 position (top panel), the S2 position (middle panel), and the B1 position (bottom panel). The profile intensity is normalized to the observations. The line emission at V$_{lsr}$ $\sim$0 \kms\ is from gas in the line of sight and unrelated to HB~3.}
    \label{fig:model_to_obs}
\end{figure}

Figure \ref{fig:model_to_obs} shows the best match of the CO(3-2) profile predicted from the shock model as compared to observations toward three positions: S1 (top panel), S2 (middle panel), and B1 (bottom panel). The best parameters from the model are listed in Table \ref{Tshockmodels}. The inferred shock velocity is 20 - 40 \kms, a density 6$\times$10$^3$ -- 2$\times$10$^4$ cm$^{-3}$ -- and the magnetic field is $\gtrsim$200$\mu$G, which is more than one order of magnitude larger than 20$\mu$G measured by radio observations toward this direction \citep{gray99}. 
\citet{gray99} estimate the magnetic field strength of 20\,$\mu G$ in average toward the W3/W4/W5/HB\,3 complexes over 40$^{\circ}$ of sky area, where much of the emission is associated with relatively diffuse, ionized gas surrounding the complex. For the regions with dense-shock tracers in HB\,3, we find the interaction is with the actual molecular cloud as opposed to the diffuse halo of the complex. For that reason,  the 20\,$\mu\,G$ represents a lower bound for the magnetic field strength of HB\,3.  It is undoubtedly much greater in the dense regions.

The magnetic field, inferred from diffuse through molecular cloud Zeeman splitting, is $B = 0.5 \times\, n^{0.65}$ $\mu$G where $n$ is a density. For a density of 10$^{4}$ cm$^{-3}$, $B$ is $\sim$200 (80 - 500) $\mu$G \citep[see Fig.\,1 of][]{crutcher10}. Our estimated strength of the magnetic field of \,$\gtrsim$200 $\mu$G does not contradict those in \cite{crutcher10}.

The preshock velocities are adopted from the narrow peak as $v_{\rm preshock}(\rm S1)\simeq 58 \rm km\ s^{-1}$, $v_{\rm preshock}(\rm S2)\simeq 40 \rm km\ s^{-1}$, $v_{\rm preshock}(\rm B1)\simeq 37 \rm km\ s^{-1}$ (see Fig. \ref{fig:model_to_obs}). Because the excitation energy of CO(3-2) is low ($E_{\rm up}=33.19$K), this line is thus easily excited long after the shock passes where gas temperature is cooled down. However, this excess is so narrow it resembles a Dirac-Delta function. Therefore, we remove its contribution in Figure \ref{fig:model_to_obs}.

For B1, the CO(3-2)  line is broad and asymmetric, and our model estimates $v_{shock}$ of 37 \kms\ and a viewing angle of 55$^{\circ}$. Here the viewing angle is defined as the angle between the line of sight and the direction vector of the shock. Note that in the one-dimensional shock model we use here, B-field direction cannot be a constraint. The shock velocity can be approximately the beginning and the end of the line profile (i.e., the width between the two zero fluxes, from -60 \kms\ to -20 \kms\ in Figure \ref{fig:model_to_obs}).  The line that is roughly symmetric (for example, like B2 in Figure \ref{hb3coh2boundary}), indicates the shock velocity to be approximately perpendicular to the line of sight. The shocked gas is blue-shifted relative to the ambient gas, as shown in Figure \ref{fig:model_to_obs}.

In contrast, for S1 and S2, the shocked gas (see Figures \ref{hb3coh2south} and \ref{fig:model_to_obs}) is redshifted relative to the ambient gas, so the shock is moving away from us, on the backside of the SNR ($\theta$ = 212$^{\circ}$ and 252$^{\circ}$, respectively), as listed in Table \ref{Tshockmodels}. The H$_2$ filament is the shell of southwest direction (see Figure \ref{hb3coh2south}), which is consistent with the angle being greater than 180$^{\circ}$. The magnetic field is largely perpendicular to the shock velocity in CJ-shock. The most narrow line profile can be for the shock penetrating the densest gas. Narrower lines at velocities far from the ambient gas can be explained by shocks into regions with a magnetic field parallel to the shock velocity, and the magnitude of the velocity shift depends upon the velocity \citep[see Figure 6 of][for comparison]{reach19}. S2 region has a higher density and a lower shock velocity than those in the S1 region.

The shocks in Regions B1-B3 and Regions S1-S2 are progressing in somewhat different directions, generally on the front and far side of the SNR, with variations due to density inhomogeneities and the detailed shape of the magnetic field. The morphology of the H$_2$ emission is highly filamentary, suggesting shock fronts in the plane of the sky, with a shock velocity perpendicular to the line of sight. Such shocks would produce lines with the limited offset in velocity with respect to the ambient gas; the lines would be wide where the magnetic field is perpendicular to the shock and narrow where the magnetic field is parallel to the shock velocity. The regions sampled by spectra B1 and B3 have large velocity offsets, and they arise from shocks that are moving toward us, on the front side of the SNR (B1 has $\theta$ = 55$^\circ$.), which is consistent with the H$_2$ thin filaments that show the arc shells in the southeastern direction (Figure \ref{hb3coh2boundary}). The presence of shocks with a wide range of angles in a small region can be explained by 3-dimensional models taking into account the wrapping of the shock front around a dense obstacle \citep{tram18}. The changes observed in line profile shape on small scales are, we suspect, due to changes in the direction of the shock relative to the magnetic field. Small-scale changes in the magnetic field are possible due to the evolution of the shocks passing through dense regions, with the magnetic field wrapping around the denser cores. The improved 3D model with applications to the millimeter CO spectra will be presented in the future.

Future high-resolution H$_2$ line spectroscopy, such as SOFIA for rotational lines and ground-based telescopes for vibrational lines will enable us to compare the shock properties -- shock kinematics, density, shock age, and the strength of magnetic field -- between H$_2$ and CO emitting gas.

%Table6
\begin{table*}
\caption[]{Application of Shock Models to CO Spectra \label{Tshockmodels}}
\begin{center}
\begin{tabular}{llrccllr}
\hline \hline
Position & Shock Velocity & Density & Magnetic Field & Viewing angle$^a$ ($\theta$) &preshock Velocity & Figure \# &  \\ 
         & V$_s$ (\kms)  & (cm$^{-3}$)          &   ($\mu$G)   & ($^{o}$) &V$_{\small preshock}$ (\kms)& \\
S1       & $\sim$37     &$\sim$6$\times$ 10$^{3}$ &  $\gtrsim$252  & $\sim$215& -58 &\ref{hb3allcospec}, \ref{hb3coh2south}, \ref{fig:model_to_obs}  \\
S2       & $\sim$22     &$\sim$2$\times$ 10$^{4}$ &  $\gtrsim$212  & $\sim$230& -40 & \ref{hb3allcospec}, \ref{hb3coh2south}, \ref{fig:model_to_obs}          \\
B1       & $\sim$37      & $\sim$6$\times$ 10$^3$ &  $\gtrsim$252  & $\sim$55 & -37 &\ref{hb3allcospec}, \ref{hb3coh2boundary}, \ref{hb3m3cospecnew}, \ref{fig:model_to_obs}  \\
\hline \hline
\end{tabular}
\end{center}
{\footnotesize $^a$The viewing angle is defined as the angle between the line of sight and the direction of shock. The angles of 215$^{\circ}$ and 230$^{\circ}$ are the shocks moving away from us, on the backside of HB\,3. The angle of 55$^{\circ}$ is the shock moving toward us, on the front side of HB\,3.}
\renewcommand{\baselinestretch}{0.9}
\end{table*}

\section{Comparison of HB\,3 with other molecular interacting SNRs}

The molecular interaction of HB\,3 is roughly half of the SNR, a position angle of 40 to 220 degrees from north, counter clock-wise. In other words, the interacting area is from northeast to east and south to southeast. H$_2$ images from \spitzer\ and WISE show that about half of the SNR is interacting with molecular clouds. Figure \ref{hb3roadmap} shows broad CO molecular lines where shocked H$_2$ is also detected. H$_2$ images critically reveal where the interaction occurs. The H$_2$ emission crucially reveals the location and energy budget of the shock passing through the region.

It is interesting that despite the half-region interaction, the radio images in the east and south still shows a relatively well-behaved symmetric shell-like morphology. In NW's direction, the radio shell is not bright, and there is no H $_2$ emission, where the medium is likely in lower density than those of H$_2$ emitting regions. This result is consistent because optical emission is bright at the west and north, where H$_2$ emission is not detected. Interestingly, the radio shell is also weak in the direction of the south-eastern shell in HB\,3 where the SNR interfaces with the H\,II region of W3.

Several SNRs of IC\,443,  W28 , W44, 3C391, W51C, HB\,21 (references in Section \ref{Sintroduction}), and G357.7+0.3 \citep{rho17} clearly show broad molecular lines (BML) which include the lines broader than 15 \kms. We can add the SNR HB\,3 as the SNR is exhibiting BML lines. Nine SNRs have been suggested to show broad molecular lines \citep{kilpatrick16}, but they are confused with many emission lines along the line of sight, and the broadening is so far only 7-10 \kms. High spectral resolution or higher transition lines of CO lines will be helpful to verify the interactions. There is another $\sim$30 SNRs which show positional coincidence between the SNR and surrounding molecular clouds \citep{jiang10,huang86}. High spatial CO observation and high transition lines are critical to verify the dynamic interactions.

HB\,3 is located at a distance similar to W28 and shows a directional density gradient, as also observed in W28. HB\,3 shows shock-wave interaction in the east and south, while W28 shows it in the eastern and northern directions. HB\,3 is an intriguing case because it shows a beautiful and clean shock front from H$_2$ imaging. A lot of single shocks are resolved since the shock is perpendicular to the line of sight in HB\,3, while the east of W28 shows shocks moving towards us and showing a mixture of multiple shock components.

\section*{Conclusion}

1. We present the detection of shocked molecular hydrogen (H$_2$) from the SNR HB\,3 (G132.7+1.3) using Palomar WIRC narrow-band imaging, the \spitzer\ GLIMPSE360, and WISE surveys, along with detections of broad CO millimeter lines using HHSMT. ROSAT mosaicked image of HB\,3 shows center-filled X-ray emission within a well-defined radio shell and HB\,3 is classified as a mixed-morphology SNR.

2. Our near-infrared narrow-filter H$_2$ 2.12 $\mu$m imaging of HB\,3 using the Palomar Hale 200 inch telescope shows that the \spitzer\ IRAC 4.5$\mu$m and WISE 4.6$\mu$m emission originates from shocked molecular hydrogen, H$_2$ emission.

3. The high-resolution images with \spitzer\ and WISE distinguish the region structures, notably the boundary of the SNR, dense molecular clouds, and W3 complex.

H$_2$ emission is observed from the east-to-south half of SNR HB\,3. The morphology exhibits a $`$butterfly' (or $`$W') shape where the south-eastern part is wrapping around the HII region of the W3 complex, indicating that the SNR interacts with the molecular cloud and the HII region. The H$_2$ emission of HB\,3 shows thin, sharp filamentary structures, and bright emission at the southern and eastern shell.

4. Broad lines of molecular CO(3-2) and CO(1-0) are detected from the southern shell of HB\,3 and at the interaction sites with molecular clouds and near the W3 complex. The widths of the broad lines are 8-20 \kms, and the profiles change on small scales, $\sim$10 to 60 arcsec. HB\,3 stands apart from other mix-morphology SNR in that we also observe narrow lines, $\sim$8 \kms (in FWHM), which might imply that individual shocks are being resolved in dense gas.

5. We apply the Paris-Durham shock model and reproduce the CO line profiles. The inferred shock velocity of 20 - 40 \kms, a density 6$\times$10$^3$ -- 2$\times$10$^4$ cm$^{-3}$, and the viewing angle indicates that H$_2$ emission in the SE is on the front side of the SNR, while the southern shell is on the backside of the SNR.

JWST with high-spatial-resolution imaging on sub-arcsec (0.1$''$ with MIRI) to two decimals ($\sim$0.05$''$ with NIRCam) scales would be able to resolve multi-shocks and 3D bending geometry that is just beyond our current observations and modeling techniques. \spitzer\ spectroscopy observations of HB~3 were not taken. The JWST spectroscopy can provide rotational transition lines of H$_2$ lines, which will enable us to probe excitation temperatures and the age of the shock of the H$_2$ emitting regions.
{\vskip 0.55truecm}
%\acknowledgements
We thank an anonymous referee for careful reading and helpful and detailed comments.
We thank GLIMPSE360 team including M. Meade, B. Babler, and R. Indebetourw for drawing our attention to the infrared emission from HB\,3. We thank A. Tappe for participating in the Palomar 200 inch observing run. This work is based on observations made with the {\it Spitzer Space Telescope}, which is operated by the Jet Propulsion Laboratory, California Institute of Technology, under NASA contract 1407. JR acknowledges support from NASA ADAP grants (NNX12AG97G and 80NSSC20K0449). THJ acknowledges funding from the National Research Foundation under the Research Career Advancement and South African Research Chair Initiative programs (SARChI), respectively. The Arizona Radio Observatory including HHSMT is part of the Steward Observatory at the University of Arizona and receives partial support from the National Science Foundation.

{\it Facilities:}
{\spitzer, WISE, Palomar Hale 200-inch (WIRC), Arizona Observatory HHSMT and 12-Meter Kitt Peak}\\ 
{\it Software:} {CLASS and GILDAS, IRTF, IDL, Python, Astropy}
\bibliography{msrefsall}

\end{document}